\documentclass{sig-alternate}

\newtheorem{problem}{Problem}
\renewcommand{\paragraph}[1]{\vspace{4pt}\noindent\textbf{#1}\xspace} 

\usepackage{bbding}
\usepackage{graphicx}
\usepackage{url}
\usepackage[usenames,dvipsnames]{xcolor}
\usepackage[parfill]{parskip}
\usepackage{caption}
\usepackage{float}
\usepackage{multirow}
\usepackage[normalem]{ulem}
\usepackage{array}
\usepackage{listings}
\usepackage{xspace}
\usepackage{pifont}
\usepackage{cleveref}

\newcommand{\xmark}{\textcolor{red}{\ding{55}}\xspace}

\newtheorem{example}{Example}

\newcommand{\eat}[1]{}

\newcommand{\sys}{CLAMShell\xspace}
\newcommand{\prob}{{\it The Crowd Labeling Problem\xspace}}

\newcolumntype{P}[1]{>{\centering\arraybackslash}p{#1}}

\begin{document}

\title{\sys: Speeding up Crowds for \\Low-latency Data Labeling}
\numberofauthors{4}
\author{
\begin{tabular}{cccc}
Daniel Haas & Jiannan Wang & Eugene Wu\large{$^{\diamondsuit}$} & Michael J. Franklin
\end{tabular}
\and 
\begin{tabular}{cc}
AMPLab, UC Berkeley & Columbia University\large{$^{\diamondsuit}$} \\
\{dhaas, jnwang, franklin\}@cs.berkeley.edu & ewu@cs.columbia.edu
\end{tabular}
}

\date{\today}

\maketitle

\begin{abstract}
Data labeling is a necessary but often slow process that impedes the development of interactive systems for modern data analysis.
Despite rising demand for manual data labeling, there is a surprising lack of work addressing its high and unpredictable latency.
In this paper, we introduce \sys, a system that speeds up crowds in order to achieve consistently low-latency data labeling.
We offer a taxonomy of the sources of labeling latency and study several large crowd-sourced labeling deployments to understand their empirical latency profiles. 
Driven by these insights, we comprehensively tackle each source of latency, both by developing novel techniques such as straggler mitigation and pool maintenance and by optimizing existing methods such as crowd retainer pools and active learning. 
We evaluate \sys{} in simulation and on live workers on Amazon's Mechanical Turk, demonstrating that our techniques can provide an order of magnitude speedup and variance reduction over existing crowdsourced labeling strategies.
\end{abstract}

\section{Introduction}

Modern data analysis is fundamentally centered around the human analyst and her ability to rapidly iterate between hypotheses and evidence.
Towards this goal numerous projects have optimized individual data analysis components
(e.g., data ingest~\cite{Abouzied:2013ig, Muhlbauer:2013fn}, data analytics~\cite{Diaconu:2013gf, Melnik:2010up, Stonebraker:2005uf, Zukowski:2012fh, Zaharia:2012ve}, visualization~\cite{Wu:2014wa, Wickham:2013ut, Liu:2013gv, Jugel:2014we} and predictive models~\cite{Ghoting:2011ux, Meng:2015tu, Jordan:2015ec, Gonzalez:2012ws}) as well as multi-stage workflows~\cite{Haas:2015a,keystoneml} to reduce end-to-end latency of data analysis.

Unfortunately, these advances continue to be hindered by the need for synchronous human effort, often in the form of manual labeling.
For example, human workers are frequently tasked to label training data (e.g., sentiment analysis, user preferences) 
for machine learning models.
Similarly, many data cleaning systems~\cite{Gokhale:2014wv, Wang:2014cf, Stonebraker:2013vl, Kandel:2011vj} rely on crowd workers to provide labels for entity resolution, value imputation, and other error mitigation algorithms.
In fact, a recent survey of software companies~\cite{crowdsurvey} found that these companies use crowd workers to complete hundreds of thousands of data cleaning tasks per day.
Such heavy reliance on manually generated data inevitably limits the speed of analysis pipelines by the latency of their crowdsourcing steps.
 
All crowd-based data labeling systems seek to reduce cost and speed while maximizing quality.
However, most research has focused only on the trade-off between quality and cost, with work on crowdsourcing routinely reporting task latencies on the order of minutes to hours to complete an average task~\cite{Bernstein:2010ha, Kittur:2008gj, Franklin:2011kr}---clearly unacceptable for user-facing data systems.

In this paper, we explicitly tackle the trade-off between cost and latency for crowd-sourced labeling tasks. 
Though there are a few existing works that explicitly aim at tackling latency, they are either tailored to specific tasks~\cite{Marcus:2012vh, Marcus:2011wf, Trushkowsky:2013kh},
targeted towards a single source of latency such as recruitment time~\cite{Bernstein:2011gh, Bigham:2010cl}, or focused on machine learning techniques (e.g., active learning) that ignore the practicalities of live crowdsourcing and may be counterproductive in terms of wall clock latency~\cite{Mozafari:2014wv}. 

In addition, predictability of overall task latency is an important consideration that has not been carefully studied. 
Depending on the numerous external factors, the quantity, quality, and speed of available workers on crowd platforms such as Amazon's Mechanical Turk (MTurk)~\cite{mturk} can fluctuate wildly~\cite{Ipeirotis:2010jo, odesk-demo} and result in individual task latencies from seconds to even days.
We argue that in order to be useful for user-facing applications, the variance of task latency must be within single-digit seconds 
before it can be embedded in interactive user-facing applications such as Data Wrangler~\cite{Kandel:2011vj}.

In this paper, we introduce \sys, a system that speeds up crowds in order to achieve consistent, low-latency data labeling. 
Rather than focus on a single algorithm or step in the data labeling lifecycle, our goal is to develop a collection of pragmatic techniques to clamp down on latency and variance during all stages of labelling.
To this end, we first perform an empirical study of the dominant sources of latency---per-task latency, 
batch-wise latency, and end-to-end overall latency.
We then systematically address each major source through three novel techniques:
\textit{Straggler mitigation} uses redundant labelers to mitigate `straggler tasks' at the end of batches, decreasing the variance of batch labeling time from minutes to fractions of seconds.
\textit{Pool maintenance} uses threshold-based eviction techniques to maintain a pool of fast, high-quality workers and decrease the average time to label each task.
\textit{Hybrid learning} combines active and passive learning to  exploit crowd pool parallelism when there are more workers available than the active learning batch size, and dynamically favors passive learning on datasets where active learning performs poorly.
Our evaluation of \sys, a system that implements these techniques on live workers, demonstrates up to $8\times$ speedups in label acquisition time and over 2 orders of magnitude reduction in variance compared to typical non-optimized deployments.
A key benefit of our work is that all of these optimizations are compatible with standard quality control algorithms such as redundancy-based voting schemes and worker quality estimation algorithms.

\section{Studying Crowd Latency}
\label{sec:latency}

In this section, we categorize the primary sources of crowdsourced microtask latency, describe existing work that addresses crowdsourcing latency,
and outline our approach towards a comprehensive solution.
We include a study of one crowd-labeling MTurk deployment that ran $\sim\!60,\!000$ tasks to label medical publication abstracts.  A full analysis of this and three other microtask deployments can be found in our technical report.

\subsection{Sources of Latency}\label{subsec:analysis}

 A multitude of factors can increase latency, from algorithm choice to worker and environmental factors.
We find that categorizing the factors based on the granularity of work provides a clear decoupling of algorithmic contributions from systems concerns.
Specifically, latency might arise from the speed of a single task, a fixed batch of tasks, or the full run of multiple batches (of possibly varying sizes).

\paragraph{Per-Task Latency} We can view the latency of a single task as a linear sequence of three phases:

\begin{enumerate}
\item{ Recruitment: } Workers do not immediately begin working on newly submitted tasks, and recruitment latency
  consists of the time until an interested crowd worker accepts a newly posted task.  
  In the medical deployment, the min, median and standard deviation statistics were $5$, $36$, and $9$ minutes, respectively.
    
\item{ Qualification and Training: } Once workers accept a task for the first time, they are often presented with tutorials 
  or qualification tasks before they are permitted to perform actual work. 
\item{ Work: } The amount of time a worker spends to complete a task can vary depending on the worker competency, 
  the time of day, fatigue, and numerous other factors~\cite{Krueger:2007ei, Ipeirotis:2010jo}.  
  Note that a single task may produce multiple labels if records are grouped into tasks (a common practice).

\end{enumerate}

\paragraph{Per-Batch Latency} 
We define the batch latency as the time for all tasks in a fixed-sized set to fully complete when sent to a crowd,
which is dependent on the {\it latency distribution} of all available workers in addition to each worker's individual variations.

For example, in the medical deployment, the median and standard deviation to complete a given HIT were $4$ and $2$ minutes, respectively, while the $90^{th}$ percentiles are upwards of $1.1$ and $3$ {\it hours}, respectively.
Although each HIT produces multiple labels, this extreme long-tail distribution is common-place on microtask platforms like MTurk, and driven by three sources:  

\begin{enumerate}
\item{ Stragglers: } The batch must block until the slowest task is completed -- up to $3$ OOM slower than the median.
\item{ Mean Pool Latency (MPL): } The expected latency depends on the MPL, which varies from $1$ to $2$ minutes.
\item{ Pool and Worker Variance: } The long-tail ultimately results in high variance within and between batches. 
The most and least consistent workers had standard deviations of $4$ minutes and $2.7$ hours, respectively.
\end{enumerate}

These sources contribute to task response times that are, in practice, slow and extremely variable.

\paragraph{Full-Run Latency} 
Rather than require crowd workers to label terabytes of data, machine learning is often used to infer labels once enough records have been labeled to train a high-quality model. 
Active learning can reduce the size of this training set, however training the model requires acquiring small batches of labels in a blocking fashion.  This induces four latency sources:

\begin{enumerate}
\item{Decision Latency:} The time to pick the next batch of tasks (e.g., uncertainty sampling for active learning)
\item{Task Count:} The number of labeling tasks, which machine learning approaches seek to reduce.
\item{Batch Size:} The batch size affects both active learning convergence as well as the amount of parallelism within a batch.
\item{Pool Size:} The number of workers completing tasks controls the maximum parallelism possible, however is often dictated by operational constraints.
\end{enumerate}

Active learning can drastically reduce the task count, but incurs increased decision latency and requires limited batch sizes to be effective.
In contrast, passive learning can leverage the parallelism of all available workers, but might require many more tasks to train a model of equivalent accuracy.
The choice ultimately depends on the labeling task, as we show empirically in Section~\ref{sec:exp-hybrid}.

\subsection{Tackling Latency}

  \begin{table}[h]\small
    \begin{tabular}{c|c|c}
    {\bf Task Latency} & {\bf Batch Latency} & {\bf Full-Run Latency} \\
    Recruitment*               & Stragglers         & Decision Time  \\
    Qual \& Training           & Mean pool latency  & Task Count*  \\
    Work*                      & Pool variance      & Batch Size  \\
                              &                    & Pool Size  \\ \end{tabular}\\
  \vspace*{-.1in}
  \caption{Classification of sources of latency in data labeling.}
  \vspace*{-.1in}
    \label{t:matrix}
  \end{table}
Table~\ref{t:matrix} summarizes the sources of latency described in the previous section, and notes (*) sources that have been addressed in the literature.  
From the table, it is clear that there is ample opportunity to improve the state of crowdsourced latency.

\paragraph{Existing Literature}
The primary work adresses recruitment time, a dominant source of task latency.
Bigham et al.~\cite{Bigham:2010cl} frequently repost tasks (among other techniques) to improve the chances of workers accepting their tasks.
However, if widely adopted, such techniques would likely exacerbate recruitment time.
Bernstein et al.~\cite{Bernstein:2011gh, Bernstein:2012va} proposed the \emph{retainer model}, which pre-recruits a pool of crowd workers (a retainer pool) and pays them to stay and be ready to accept tasks.
In settings where tasks are streaming or come in batches, this model can effectively eliminate recruitment time at a small cost.
In our work, we build on top of the retainer model.

Work time has been reduced by re-designing task interfaces~\cite{Marcus:2012vh}.
For example, Marcus et al.~\cite{Marcus:2011wf} study join interfaces for images, and design interface batching techniques that let workers complete up to 9 pair-wise comparisons in the same time as a single pair-wise comparison task.  
However, these approaches are task specific, so \sys views them as complementary to its general task optimization framework and does not explicitly address them.

Finally, algorithmic analysis and machine learning have been used to reduce task count.
The former focus on efficient algorithms for specific operations (e.g., entity resolution~\cite{crowder}, counting~\cite{Marcus:2012vh}, or information retrieval~\cite{DasSarma:2014kz,Parameswaran:2013td}).  
These focus on full-run latency, and could leverage \sys's per-task, per-batch, and machine learning techniques.

The latter trains models using data from completed tasks until the prediction quality exceeds a user-defined threshold, and then is used to predict the remaining responses. 
In this setting, active learning~\cite{Cohn:1996to} is a commonly used method~\cite{Mozafari:2014wv, Gokhale:2014wv}.
Given unlabeled data, active learning iteratively uses a point selection algorithm to pick a small set of informative points to acquire labels for,  and incorporates the new labels into its model.
The algorithm continues until the model accuracy (e.g., cross-validation) converges.
Active learning is indispensible when there are more items than can be practically labeled, and can be used in conjunction with algorithmic approaches that rely on the labels~\cite{DasSarma:2014kz, crowder}.

Despite reducing the task count, active learning may counter-intuitively {\it increase} the overall latency by  
constraining the parallelism due to its batch size limitations.  
Its convergence properties have only been proved when the batch size is 1.
and  larger batch sizes (e.g., $10$)  have only been tested empirically.
When the number of workers significantly exceeds the batch size, active learning can be much slower than 
labeling as many random tasks in parallel as possible and using a passive learner.

\paragraph{Towards a Comprehensive Solution}
The core problem is a trade-off between cost and latency:

\begin{problem}[\prob]\label{activeclean}\sloppy
A user wants to label $N$ items using a pool of $p$ workers at an accuracy level of $\alpha$ (e.g., $\alpha\%$ of all items are labeled correctly).
Minimize the metric $\frac{1}{\beta l + (1-\beta) c}$
where $l$ is the latency to label the items, 
$c$ is the total used cost, and
$\beta$ is a user-specified parameter expressing a preference for speed versus cost.
\end{problem}

To this end, we systematically tackle the primary sources of latency (Table~\ref{t:matrix})
in a general purpose labeling system:

\begin{enumerate}
\item{Task Latency:} \sys{} addresses task latency by adopting retainer pools to reduce recruitment costs. 
	\sys{} automatically maintains the pool size at $p$ as workers abandon the pool, and provides guidance about how the cost and latency will be affected by changing $p$.  
	In addition, \sys{} trains and verifies worker qualifications as part of recruitment, ensuring that every worker in the pool is immediately available to provide useful work when new tasks arrive. 
	
\item{Batch Latency:} 
{\it Straggler mitigation} uses worker redundancy on slow tasks to compensate for long-tail latencies. 
{\it Pool maintenance} selectively replaces pool workers to progressively shift and tighten the latency distribution towards faster responses.
Together, they eliminate straggler effects, reduce mean pool latencies over time, and significantly reduce batch variance.
	
\item{Full-Run Latency:}
  \sys uses a hybrid strategy that allocates subsets of the worker pool to active and passive learning.
  In addition, \sys pipelines the expensive model retraining and uncertainty sampling steps with crowd labeling to
  eliminate decision latency at the cost of slightly stale model results.

\end{enumerate}

Note that \sys{} does not explicitly address {\it work time}, nor {\it pool size}:
work time is often specific to the task interface, which we view as an orthogonal interface optimization problem,
and pool size is a parameter to \prob{} and typically set by operational constraints. Instead, the following text focuses on the other sources of latency listed in Table~\ref{t:matrix}.

\section{The \MakeUppercase\sys{} System}
\label{sec:arch}

In this section, we present an overview of \sys, a system for
fast label acquisition.

\begin{figure}[t]
\centering
 \includegraphics[width=.8\columnwidth]{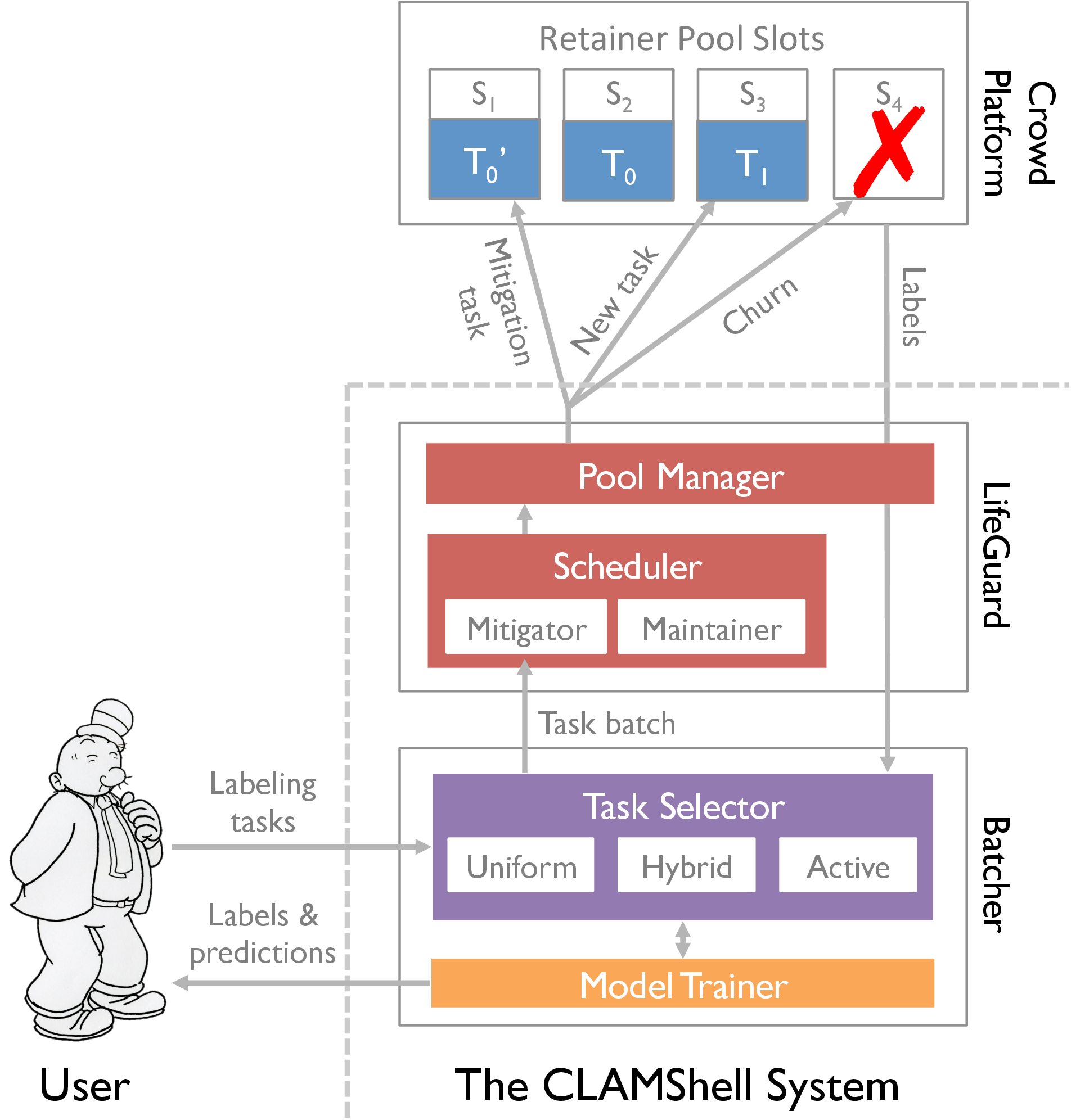}
 \vspace{-.2cm}
 \caption{\sys architecture diagram.}
 \label{f:arch}
 \vspace{-.3cm}
\end{figure}

The \sys architecture is illustrated in Figure~\ref{f:arch}. 
The user submits a set or stream of labeling tasks to the {\it Batcher} and uses the {\it Task selector} (Section~\ref{sec:hybrid-learning}) to pick $B$ incomplete tasks to process in the current iteration. The tasks are selected via uncertainty sampling using the most recently trained model to pick tasks that benefit active learning, and random sampling to pick tasks for 
passive learning.
The resulting batch is sent to {\it LifeGuard}, which schedules tasks within the batch to be sent to the {\it Crowd Platform}.  
This level of indirection is necessary when the batch size exceeds the size of the retainer pool, and so the {\it Mitigator}
can control redundancy when there are slow tasks.

The {\it Crowd Platform} holds a set of slots ($S_1\ldots S_4$) in the current retainer pool.
Each slot corresponds to a persistent {\it retainer task} that a crowd worker has accepted, and may be empty (e.g., $S_4$) or contain a task (e.g., $T_0$).
The {\it Scheduler} immediately sends new tasks to available slots (e.g., $S_3$).
If all tasks have been sent, then the {\it Mitigator} sends duplicate (mitigation) tasks for slow, incomplete tasks (e.g., $S_1$).
If a slot is consistently performing slowly, the {\it Maintainer} may recruit and train a worker for a replacement slot in the background, and evict the slot (\xmark in $S_4$) when the new one is available.

Completed labels are sent directly to the {\it Batcher}, which retrains the machine learning model. 
The {\it Task Selector} uses different sampling algorithms such as uniform sampling, active learning-based uncertainty sampling,
or our hybrid sampler, to pick the next batch of tasks.
During the entire process, the user receives the completed labels, and is able to query the currently trained model
for new predictions.
Next, we show an example to describe the use of \sys in practice.

\begin{example}
Imagine a news outlet is covering a live political debate, and wants to monitor and visualize the public's reaction to candidates' comments on hot-button issues by analyzing the sentiment of related tweets. 
Because automated sentiment analysis techniques on tweets are often inadequate~\cite{Agarwal:2011tc}, the company asks a crowd to label tweets as ``positive", ``negative", or ``neutral". 
If the system suffered from high crowd latency, the sentiment visualization would be unable to keep up with the changes in public opinion as the debate proceeded, rendering the tool unhelpful.

\sys can be used to address this issue with both per-batch and full-run optimizations.
The per-batch optimizations, including straggler mitigation and pool maintenance techniques, are designed to reduce the time that is required to label a batch of tasks using crowds, e.g., asking for crowd labels for a batch of ten tweets.  
Once the company has enough labeled data, they hope to switch to an automated process in the long term.  
The full-run optimizations, including hybrid learning, are designed to reduce the number of iterations that a learning model needs to converge, i.e., the total number of batches that we need to ask crowds to label.
\end{example}

\section{Per-batch Latency Optimization}
\label{sec:per-batch}

Per-batch optimizations aim to reduce the latency for a single batch of labeling tasks --- the {\it Batcher} sends a 
batch of tasks to a pool of workers, and waits for the batch of work to complete.  
In this model, the dominant costs are due to the variability of worker latencies within the pool, 
as well as the variability within the tasks that a single worker performs.

\begin{figure}[h]
\centering
\vspace{-.2cm}
 \includegraphics[width=.8\columnwidth]{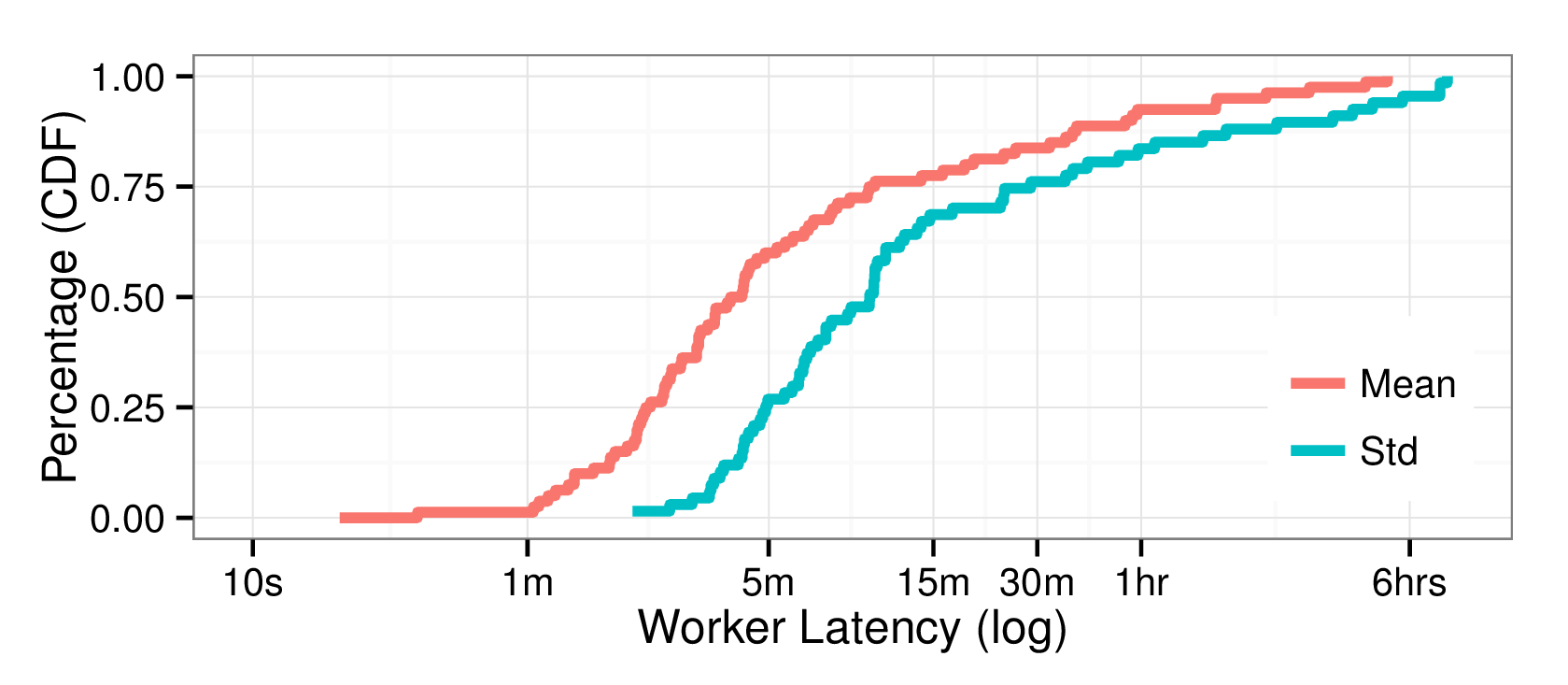}
 \vspace{-.4cm}
 \caption{Distribution of worker latencies.}
 \label{f:dist_kraska_worker}
 \vspace{-.3cm}
\end{figure}

For example, Figure~\ref{f:dist_kraska_worker} depicts per-worker means and standard deviations of latency from the medical deployment as CDFs. We can see that
average worker speeds are spread out from tens of seconds to hours. 
In addition, even workers who are very fast on average ($\sim\!1$ minute)
can take as long as an hour or more to complete some tasks.
This variation is bad for per-batch latency because the batch must block until all of its tasks are complete.

So in order to reduce per-batch latency, a system must reduce both the mean of the latency distribution (workers who are slow on average) and its variance (workers who are inconsistent).
This section describes the mechanisms for each of these approaches respectively, along with mathematical models and
simulation results.  Section~\ref{sec:eval} evaluates these strategies on a live deployment on MTurk.

Throughout the following sections, we reference experiments run in simulation. 
The simulator setup is described in Section~\ref{sec:setup}, but due to space constraints detailed analysis of the results is omitted and can be found in our technical report.

\subsection{Straggler Mitigation: Reducing Variance}
\label{sec:stragglers}

In cluster computing frameworks such as Hadoop~\cite{hadoop} or Spark~\cite{Zaharia:2012ve} where the presence of straggler tasks in a stage (e.g., reduce stage of MapReduce~\cite{Dean:2008fi}) can delay downstream computation, replicating the slow tasks~\cite{Ananthanarayanan:2013vg, Dean:2008fi, Ananthanarayanan:2010tx, Zaharia:2008um} via speculative execution or task cloning~\cite{Ananthanarayanan:2013vg}
is an effective counter-measure.  

We take a similar replication-based approach to human stragglers in our crowd pool.
We call a worker {\it active} if she is currently working on a task, and {\it available} otherwise.  
Similarly, a task is either {\it active}, {\it complete}, or {\it unassigned}.
By default, \sys routes only unassigned tasks to available workers until all tasks are complete.
Once all tasks are active or complete, available workers must wait until the next batch to receive a task.
With straggler mitigation, in contrast, such workers are immediately assigned active tasks, creating duplicate assignments of those tasks.
\sys returns the first completed assignment of a task to the user and immediately reassigns all other workers still working on that task to a new unassigned or active task (though it pays them for their partial work on the old task regardless).
The effect of straggler mitigation is that when an inconsistent worker takes a long time to complete a task, the system hides that latency by sending the task to other, faster workers.
As a result, the fastest workers complete the majority of the tasks and earn money commensurate with their speed.
For example, in the medical deployment, the fastest worker ($\mu=28.5$ seconds) could complete, on average, $8\times$ as many tasks as the median worker ($\mu=4$ minutes).

\paragraph{Simulation.} A natural question arises when performing straggler mitigation: which task should be assigned to an available worker?
We ran simulation experiments testing several straggler routing algorithms, including routing to the longest-running active task, to a random task, to the task with fewest active workers, or to the task known by an oracle to complete the slowest.

To our surprise, the selection algorithm didn't affect end-to-end latency, and random performed as fast as
the oracle solution because the fast workers complete tasks so quickly that they complete 
almost all of the tasks in the batch anyways.

A second question is: at what batch sizes is straggler mitigation effective?  
We study this in simulation by varying the pool size to batch size ratio $R=\frac{N_{pool}}{N_{batch}}$ 
using the random selection algorithm and different pool sizes.
The benefit of straggler mitigation comes from its ability to remove the overhead of slow workers at the end of a batch of tasks.
When $R$ is higher, each batch gains the full benefit of straggler mitigation and completes at the speed of the fastest workers, 
however the number of tasks completed in each batch is lower.
Conversely, with a small ratio, workers spend most of their time working on unassigned tasks, and the impact of straggler mitigation is lessened.

\paragraph{Impact on Crowdsourcing Systems.} Straggler mitigation is a general technique that does not affect the programming interface of the system it is applied to.
It can therefore be used easily in conjunction with any existing crowdsourcing system that processes batches of microtasks.
One important benefit of hiding the variance in worker latencies is that task completion times become much more predictable.
This characteristic is vital to the development of declarative crowd systems such as crowdsourced query processors, because optimizers need to be able to accurately estimate the cost of executing a declarative crowd workflow.

\paragraph{Working with Quality Control.}  Straggler mitigation is reminiscent of redundancy-based quality control algorithms such as~\cite{Ipeirotis:2010do} or~\cite{Karger:2011wh} that use votes from multiple workers to better estimate the true answer.
However, straggler mitigation stops as soon as it has a single answer in order to return as quickly as possible.
A na\"{i}ve combination of straggler mitigation and quality control might be inefficient.
For example, duplicating a task for straggler mitigation that requires 3 votes for quality control would create 6 assignments, whereas perhaps only 4 or 5 are necessary to get 3 answers without any straggling tasks.
In order to avoid this effect, \sys decouples straggler mitigation assignments from quality control assignments. 
That is, a quality-controlled task is marked as \textit{active} until it has received (say) 3 answers, and straggler mitigation assigns only single available workers to the task at a time to eliminate stragglers.
In simulation, we find that this optimization can provide up to 30\% per-batch latency improvement in settings where stragglers are much slower than average workers and most of the pool is composed of fast workers.

\subsection{Pool Maintenance: Better Mean Latency}
\label{sec:churn}

Straggler mitigation reduces the variance of task latencies, but if many workers in the labeling pool are slow on average, variance reduction will be ineffective at reducing per-batch latency.
To improve the average speed of the pool over time, \sys uses \textit{pool maintenance}, a technique that continuously replaces slow workers in order to converge to a pool of mostly fast workers.
Because a fast pool will label each task more quickly, pool maintenance reduces per-batch labeling latency over time.

Our maintenance algorithm takes as input a latency threshold $P\!M_\ell$, and continuously releases workers slower than the threshold asynchronously as labeling proceeds.
To do so, it computes an empirical latency for each pool worker based on the worker's completed tasks
and flags the worker as a candidate for removal if his latency is significantly above $P\!M_\ell$ (determined using a one-sided significance test).

Instead of removing a slower worker before recruiting a replacement, \sys continuously recruits and trains workers in the background in order to maintain a reserve of new workers.
Although this might seem costly, pipelining recruitment means that pool maintenance can proceed without blocking on worker recruitment, and we find empirically that the latency savings of pool maintenance translate to cost savings that overwhelm the cost of background recruitment (Section~\ref{sec:churn-eval}).
The removed worker is paid for their active job (if any), and informed that there are no more tasks available for the experimental run.
They are not blacklisted, so that future experiments are not biased.

\paragraph{Pool speed convergence.} The following model demonstrates the mean latency to which a maintained pool will eventually converge. 
Assume a population of workers with mean latencies $\mu_i$ following some global distribution $\mathcal{W}$ having mean $\Gamma$, and sample an initial pool $\mathcal{P}_0 \subset \mathcal{W}$ uniformly at random from $\mathcal{W}$. 
Let $P\!M_\ell$ be a latency threshold splitting the distribution $\mathcal{W}$ into two parts, with probability densities $q$ and $1-q$ above and below $P\!M_\ell$ respectively. 
Further, let $\mu_f$ be the mean latency among fast workers having $\mu_i < P\!M_\ell$, and let $\mu_s$ be the mean latency among slow workers having $\mu_i > P\!M_\ell$.

Then our initial pool has a mean latency $\mathbb{E}[\mu_i] = (1-q)\mu_f + q\mu_s$. If at each maintenance step, we remove all slow workers having $\mu_i > P\!M_\ell$ and replace them with workers drawn randomly from $\mathcal{W}$, and letting $\mathcal{P}_i$ be the pool after $i$ steps, we see that $\mathcal{P}_1$ has mean latency $\mathbb{E}[\mu_i] = (1-q)\mu_f + (q(1-q)\mu_f + q^2\mu_s)$, and in general $\mathcal{P}_n$ has mean latency:
\begin{align*}
\mathbb{E}[\mu_i] &=(\sum_{i=0}^n q^i)(1-q)\mu_f + q^{n+1}\mu_s\\
&=(1-q^{n+1})\mu_f + q^{n+1}\mu_s.
 \end{align*}%
We observe that $\lim_{n\to\infty}\mathbb{E}[\mu_i] = \mu_f$, that is, the pool converges to the mean latency of all workers below $P\!M_\ell$.
This implies that it is desirable to set $P\!M_\ell$ as low as possible: in practice, setting the threshold too low leads to thrashing, as we show in section~\ref{sec:churn-eval}.

\paragraph{Simulation.} We simulated how pool maintenance affects batch latency with respect to the task to pool size ratio $R$ using a latency threshold $P\!M_\ell$ of one standard deviation below the mean. 
After each batch, we replace all workers slower than $P\!M_\ell$ with new samples from the worker distribution.
With pool maintenance, the batch latency falls quickly, nearly halving in just 15 to 20 batches. 
When there are many more tasks than pool workers, the effect becomes less pronounced, because there are enough tasks that slow workers who only complete a small fraction of tasks do not impact the per-batch latency.

To better understand how the distribution of mean worker latency is changing over time, we simulate the mean pool latency (MPL) of the worker pool over time with and without maintenance, and compare the MPL to the mathematical model's predictions.
With maintenance, the pool's MPL converges quickly to the model's predicted asymptote, following the model closely across pool-size to task ratios $R$.

\paragraph{Latency Threshold.} The pool maintenance latency threshold determines which workers are slow and should be removed from the pool. 
To pick a good threshold, we can observe the empirical distribution of all workers ever seen, and estimate the threshold as $k$ standard deviations below the mean. 
The goal is to find a threshold low enough to decrease average pool latency by releasing slow workers, but high enough to avoid discarding the fastest workers from the pool. 
In Section~\ref{sec:churn-eval}, we vary the threshold and find that it has significant impact on the benefits of pool maintenance.

\paragraph{Extensions.} As described, pool maintenance is focused only on reducing the mean latency of the pool. However, it can be easily extended to optimize for other criteria by choosing an objective function other than worker speed. For example, we could maintain a pool using quality (estimated using, e.g., inter-worker agreement~\cite{CallisonBurch:2009ui}) to converge to a high-quality pool, use a weighted average to trade off quality and speed, or minimize another metric such as worker variance. Ramesh et al.~\cite{Ramesh:2012ta} take a similar approach to identifying high-quality workers, though they use an oracle to determine accuracy and evaluate their technique only in simulation.

\subsection{Combing Per-Batch Techniques}
\label{sec:stragchurn}
Both straggler mitigation and pool maintenance deal with tail latencies --- maintenance detects and removes workers whose average speeds are outliers, and straggler mitigation hides individual workers' outlier tasks.
From our initial live experiments, we were surprised to find that naively combining the two techniques together resulted in zero or even negative gains as compared to straggler mitigation alone.
For example, the number of workers replaced in each batch was reduced from $\sim30$ to less than $5$ despite similar worker distributions.

The reason is that straggler mitigation prevents high latency tasks by terminating the slower replicas.
A consequence of this technique is the lack of high latency tasks, which artifically skews every worker's completion times towards the latency of the fastest workers, and makes directly measuring true worker latency infeasible.
In response, we developed a simple model called \texttt{TermEst} to estimate the average latencies of terminated tasks based on the number of times a worker's task is terminated.

We assume the worker pool is represented by two workers --- a slow worker $w_s$ and a fast worker $w_f$ that each uses a true latency of $l_{s,j}$ and $l_{f,j}$ to complete task $t_j$ --
and our goal is to estimate the latency of $w_s$' terminated tasks.
Let $w_s$ start $N$ tasks $T_{all} = \{t_1, \ldots, t_N\}$, where $T_t \subseteq T_{all}$ are terminated, and $T_c = T_{all} - T_t$ are completed.
Let $l_{k,T} = \frac{1}{N} \sum_{t_i \in T} l_{k,i}$ be the average latency for $w_k$ to complete a random task in $T$, and let $l_k$ be $w_k$'s true mean latency.
Assuming that $w_f$ can start working on $t_j$ at any time after $w_s$ with uniform probability,
the probability that $w_f$ starts early enough to finish and cause $w_s$ to terminate is $\frac{l_{s,j} - l_{f,j}}{l_{s,j}}$.
Thus, $w_s$ is expected to be terminated $N_t$ times after starting $N$ tasks:
{\small
$$ \sum_{t_i \in T_t} \frac{l_{s,i} - l_{f,i}}{l_{s,i}} \approx \frac{l_{s,T_t} - l_{f}}{l_{s,T_t}}\times N = T $$}Rearranging the terms, we can estimate $l_{s,T_t}$, where $N_c = N-N_t$:
{\small
$$ l_{s,T_t} = \frac{l_{f}\times N}{N_c}$$}We then add a smoothing term $alpha$ to $N$ in order to compensate for the lack of latency evidence when $N$ is small 
and avoid divide-by-zero errors when all of a worker's tasks are terminated ($N=T$).  
In practice, we estimate $l_{f}$ as the empirical mean of the workers that caused any of $w_s$' past jobs to terminate:
{\small
$$ l_{s,T_t} = \frac{l_{f} (N + \alpha)}{N_c + \alpha}$$}Finally, to estimate the overall latency of $w_s$ by taking the 
the weighted average of $l_{s,T_t}$ and the empircal mean latency of the tasks $w_s$ is able to complete, $l_{s,T_c}$:
{\small
$$l_s = \frac{N_t}{N}\times l_{s,T_t} + \frac{N_c}{N}\times l_{s,T_c}$$}Note that our formulation is equivalent to modifying the latency threshold on a \emph{per worker} basis.
Thus, while changing the global latency threshold is important for setting a worker replacement rate, 
this adjustment replaces workers who are frequently terminated.

\section{Full-Run Latency Optimization}
\label{sec:cross-batch}

In order to eliminate the need to manually label all points in a potentially large set, \sys{} acquires labels for only as many points as needed to train a predictive model of sufficient quality, then uses that model to impute labels for all remaining points. 
As described in Section~\ref{sec:latency}, there are many factors that influence the latency of the labeling process. 
Relying on learning greatly decreases the \textit{task count} necessary to label the entire dataset, but has implications for the \textit{decision latency} and \textit{batch size} involved.
In particular, \sys{} uses active learning techniques to reduce the task count even further, but trades this improvement for increased decision latency (the learner must choose which points to label next) and decreased batch size (active learning is inherently iterative and cannot label as many points in parallel).

In this section, we describe how \sys{} ameliorates the drawbacks of active learning for low-latency labeling. 
We introduce \textit{hybrid learning}, a novel technique which combines active and passive learning to maximize pool parallelism and hide the inherent limits of active learning batch size.
We also describe how \sys{} leverages existing techniques to set an effective batch size for active learning and uses asynchronous model retraining to hide active learning's decision latency.

\subsection{Hybrid learning}
\label{sec:hybrid-learning}

Active learning uses the current trained model to decide which points to label in the point selection phase, reducing the number of points needing labels in order to train a high-quality model. 
In practice, however, there are two major challenges to active learning at low latency.
First, at each iteration, active learning has a limited batch size---setting the batch size too high can cause the model to converge even more slowly than passive learning~\cite{Settles:2010vo}.
This limits the wall-clock speed at which active learning can proceed.
Second, when labeling work is challenging, it will be hard to train a good model. 
As a result, the current trained model may misguide the point selection phase, and active learning may perform poorly, perhaps even worse than passive learning. 
On the other hand, passive learning that trains a model using a randomly sampled data points can proceed as fast as the crowd can label, but it will waste human effort for easy labeling work.

\sloppy

To address these issues, we propose \emph{hybrid learning} in \sys, with the basic idea of maintaining the best traits of both passive and active learning, allowing for fast model convergence on both easy and hard data labeling work. 
Hybrid learning simultaneously acquires labels using the active selection strategy and random sampling, maximizing crowd worker parallelism and compensating for datasets where active learning alone would perform poorly.
As a result, label acquisition can proceed at high speed in spite of a low active learning batch size.

\paragraph{Point Selection.} Once a batch size has been selected for active learning (Section~\ref{sec:batch-size}, below), hybrid learning attempts to maximize crowd worker parallelism by ensuring that each worker in the pool has at least one point to label.
That is, given a batch size $k$ and a pool size $p$, hybrid learning uses the active selection criterion to choose $k$ points for labeling, then randomly selects $\max(0, p-k)$ points for passive labeling.
Because \sys{} caches all previously labeled points, if the points chosen for active or passive labeling overlap, their labels are read from the cache and additional points are selected for labeling.

\fussy 

\paragraph{Model Retraining.} Once a new batch of points has been labeled, hybrid learning retrains a model on all previously observed labels.
These points come from two sampling distributions: uncertain sampling (active learning) and random sampling (passive learning).
Currently, \sys retrains the model on the union of these points without distinguishing their difference, though it does weight points based on the active-to-passive ratio (i.e., $\frac{k}{p}$).
If users provide hints to \sys about how hard their labeling work is (e.g., very difficult), \sys can adjust these weights accordingly.
We leave the exploration of optimal re-weighting schemes for future work.

\subsection{Active learning batch size}
\label{sec:batch-size}
Because the speed of active learning is constrained by the size of its batches, setting a good batch size is important for fast convergence.
Too small, and training will be slow because it will take a long time to label all the points.
Too large, and training will be slow because each batch contains less useful points, slowing down convergence to a good model (or even converging to a bad one!).
The literature provides no guidance on an appropriate batch size for batch-mode active learning, assuming that that the batch size is chosen by the user in advance.
Chakraborty et. al~\cite{Chakraborty:2015fh} offer an active learning technique that dynamically sets the batch size, but it is not generic across learners and requires knowledge of the labeling time for each instance.
We experimented extensively with the active learning batch size, and found that once batch size was within a reasonable range (10-40), there was no significant correlation between batch size and convergence rates on any single dataset, let alone across datasets.

As a result, we rely on empirical results from our hybrid learning experiments (Section~\ref{sec:exp-hybrid}) to set an active learning batch size that works well with our hybrid strategy.
Those experiments show that the fraction of the pool $r = \frac{k}{p}$ allocated to active learning has a significant impact on the convergence of the learner, and that $r=0.5$ is a reasonable value for multiple datasets.
In our end-to-end experiments, we set $k = 0.5p$ accordingly.

\subsection{Active learning decision latency}
\label{sec:decision-latency}
The time taken by the active learner to retrain a model and select a new batch of points after the previous batch has been labeled has a significant impact on full-run latency, because the labeling process blocks until the learner is ready with the next batch.
To mitigate this latency (which is not an issue for passive learning), \sys{} uses two known techniques.

First, rather than consider all unlabeled points for selection in the next batch, we consider only a uniform random sample of the points. 
This has been shown to have little impact on active learning convergence, and offers significant performance improvements: the point selection time is linear in the sample size, not the size of the entire unlabeled dataset, which might include millions of examples.

Second, rather than performing retraining and selection synchronously at the end of each batch, \sys{} continually retrains models asynchronously on the latest available points.
A new batch of points is selected based on each new model, so at any point in time there is an available model and an available selection of points for the next batch.
When each batch of points completes the labeling process, the next batch is selected based on the most recently computed model.
This trades off decision latency for staleness of points to be selected, and empirically we find that it does not significantly impact model convergence.

\newcommand{\straggler}{{\textsf{straggler}\xspace}}
\newcommand{\pool}{{\textsf{pool}\xspace}}
\newcommand{\hybrid}{{\textsf{hybrid}\xspace}}

\subsection{Putting it all together}
\label{sec:combined-techniques}
\sys is powered by three techniques: Straggler Mitigation (\straggler), Retainer Pool Maintenance (\pool), and Hybrid Learning (\hybrid). Table~\ref{t:cmp} summarizes their impact on system performance across four axes: (1) Can they improve the mean latency of labeling? (2) Can they mitigate the variance of individual workers' labeling latency? (3) Do they require additional cost to use? (4) Are they general or restricted to a certain labeling setting?

\def\arraystretch{1.1}
\begin{table}[t]\small
    \begin{tabular}{l|c|c|c|c}
    {\bf  \sys } &  \multicolumn{2}{c|}{\bf Latency} & \multirow{2}{*}{\bf Cost} & \multirow{2}{*}{\bf General} \\ 
     {\bf  Techniques}     & {\bf Mean}  & {\bf Variance} & & \\ \hline
      \straggler   & Yes & Yes  & Increase  & Yes\\ 
      \pool        & Yes & Yes  & No Change & Yes \\ 
      \hybrid      & Yes & No  & Increase  & AL  \end{tabular}
      \vspace{-.3cm}
    \caption{\small \sys techniques (AL: Active Learning).}
    \label{t:cmp}
\end{table}
\vspace{-.4cm}

\section{Evaluation}
\label{sec:eval}

In this section, we evaluate \sys{} both in simulation and on live crowd workers on MTurk in order to show that it enables data labeling to proceed at interactive speeds. We first evaluate each technique in isolation, then provide end-to-end experiments demonstrating the total time it takes to label unlabeled datasets. Table~\ref{tab:exp-params} summarizes important parameters varied in the experiments.

\subsection{Experimental Setup}
\label{sec:setup}

\begin{table}[t]
\small
\begin{tabular}{c | l }
{\bf Param}           & {\bf Description} \\ \hline
$P\!M_\ell$                 & Latency threshold for pool maintenance \\
$S\!M$                  & Straggler mitigation: on ($S\!M$), off ($N\!oS\!M$). \\
$N_p$               & Number of workers in the retainer pool.\\
$N_g$                 & Task complexity: \# records grouped a HIT. \\
                      & Simple (1), Medium (5), Complex (10 records)\\
$R$                   & Pool-batch ratio. \\
$Alg$                 & Learning algorithm: active ($AL$), passive ($PL$),\\
                      & hybrid learning ($HL$), or none ($NL$) \\
\end{tabular}
\vspace{-0.4cm}
\caption{Experimental Parameters}
\label{tab:exp-params}
\vspace{-0.5cm}
\end{table}

\paragraph{Simulator.} The simulated experiments described in the previous text and the following evaluation are run on a python simulator that models a retainer-pool crowd data labeler and implements uncertainty sampling on top of scikit-learn's model training~\cite{scikit-learn}. 
To simulate crowd workers, we use traces from the medical deployment described in Section~\ref{subsec:analysis}.
From each trace, we measure each worker's mean labeling latency $\mu_i$, variance in labeling latency $\sigma_i^2$, and mean accuracy $\lambda_i$.
We then generate a worker's latency on an assigned labeling task by drawing a sample i.i.d from $\mathcal{N}(\mu_i, \sigma_i^2)$, and generate the label itself by returning the correct label with probability $\lambda_i$ and the incorrect label with probability $1 - \lambda_i$.
Using these worker pools, the simulator can model recruitment (adding random workers to the pool), pool maintenance (releasing workers with high observed $\mu_i$ from the pool), straggler mitigation (assigning multiple simulated workers to the same task and returning the minimum of the sampled latencies), and active learning (using simulated workers to label batches of points and measuring the latency of the whole batch).

\paragraph{Live Experiments.} The live experiments discussed below run on a custom implementation of the retainer model for MTurk.
Recruitment occurs by repeatedly re-posting recruitment tasks every $3$ minutes to MTurk until the desired number of workers have joined the pool.
Workers are paid \$.05 / minute to wait for available work once they join a pool, and \$.02 / record to perform the work once it becomes available.
MTurk tasks require a minimum qualification of 85\% worker approval to join a pool.
Experiments in these pools were run at multiple times of day on both weekdays and weekends.
In contrast with prior work, we found that results were remarkably consistent across these parameters when using our latency mitigation techniques.
This may be the result of our relatively strict qualification requirement, or may reflect more systemic changes in the MTurk marketplace.
Following the retainer pool model, we assume recruitment time is amortized across batches and measure latency from the moment the first task is sent to the pool, rather than from the beginning of the recruitment process.
Overall, we collected timing results for nearly 250,000 individual task assignments over the span of several weeks.

\paragraph{Datasets.} The active learning tasks run in this evaluation are all classification tasks, based on publicly available datasets.
The MNIST dataset~\cite{Lecun:1998a} contains 70,000 black and white images of handwritten digits, and the multi-class classification task is to detect which digit an image represents.
We used raw pixel values as features, leading to 784 features per image.
The CIFAR-10 dataset~\cite{Krizhevsky:2009a} contains 60,000 color images of various objects, and the classification task is to identify the category of the primary object in each image.
In order to make the learning task simpler, we limited the topic categories to two: ``Birds" and ``Airplanes".
We used raw pixel values as features, generating 3072 features per image.
In addition to the real datasets, which have concrete labeling tasks that we can send to human workers, we also generate datasets of varying difficulty to illustrate the relationship between problem hardness and the performance of our techniques. These datasets are generated with the scikit-learn data generator, which builds classification problems following an adaptation of the algorithm from~\cite{Guyon:2003a}.

\subsection{Pool Maintenance}
\label{sec:churn-eval}
In this section, we evaluate the effects of pool maintenance on batch time.
The experiments execute $500$ tasks that label MNIST digit images. 
We compare tasks of varying complexity (Simple, Medium, Complex) that use $N_g=1,5,\text{or }10$ MNIST images, respectively.
The latency threshold is set to $P\!M_\ell = 8$ and $P\!M_\ell= \infty$ (no maintenance).

\begin{figure}[h]
\vspace*{-.15in}
\begin{center}
\includegraphics[width=\columnwidth]{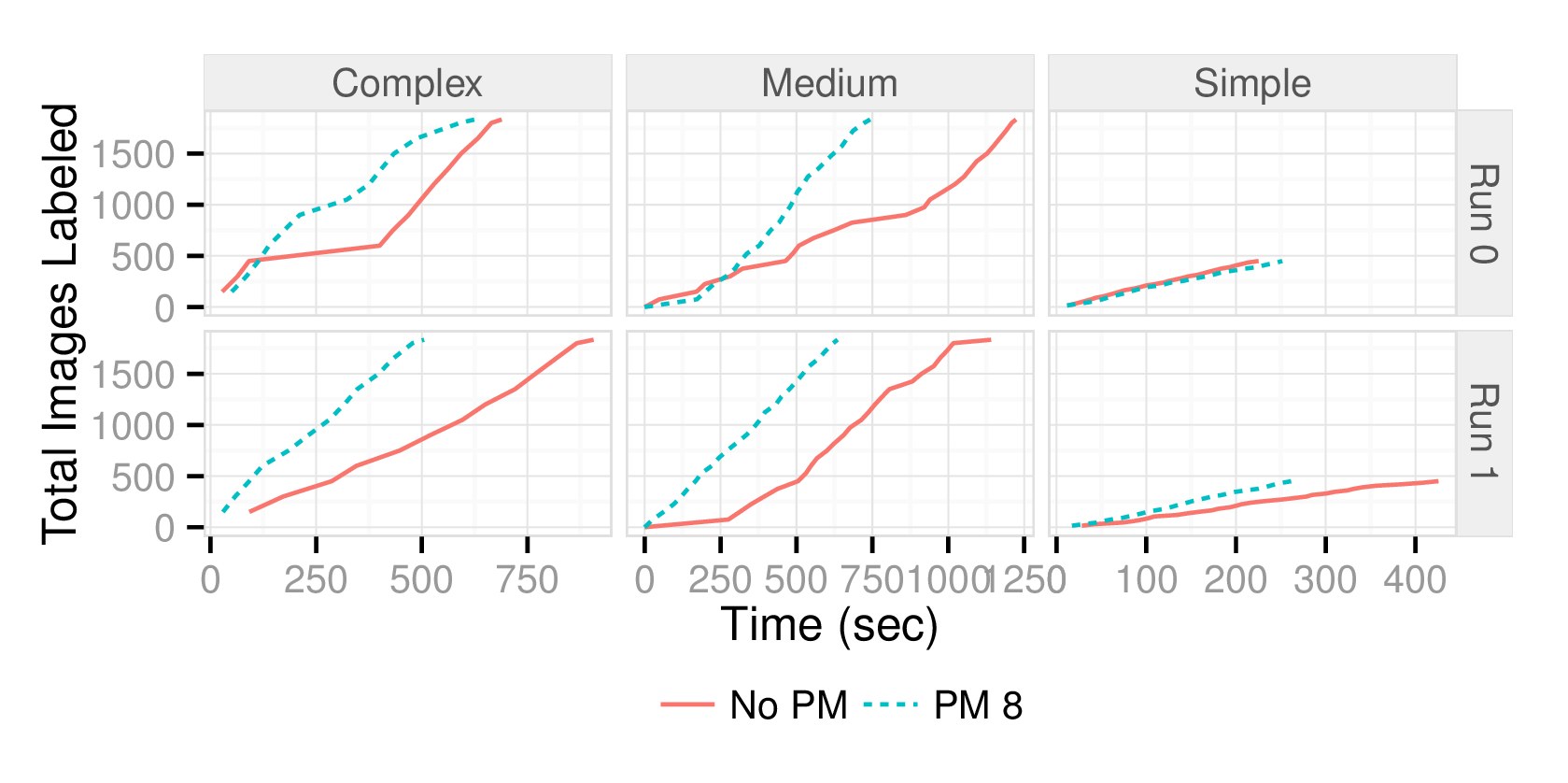}
\end{center}
\vspace*{-.2in}
\caption{\# points labeled over time.}
\label{fig:churn-npts-v-time}
\vspace*{-.1in}
\end{figure}

Figure~\ref{fig:churn-npts-v-time} is an overview of the total number of labeled points ($N_g \times N_{tasks}$) over time for each configuration.  
The slope of each curve describes the speed of task completion, where a flat curve denotes stragglers that take a very long time to complete a task.  
We find that task completion for simple tasks is uniformly fast, so pool maintenance provides little additional benefit;
however, more complex tasks are affected by outliers, and maintenance's ability to cull slow workers helps reduce the presence of very long tasks.

\begin{figure}[t]
\vspace*{-.1in}
\begin{center}
\includegraphics[width=\columnwidth]{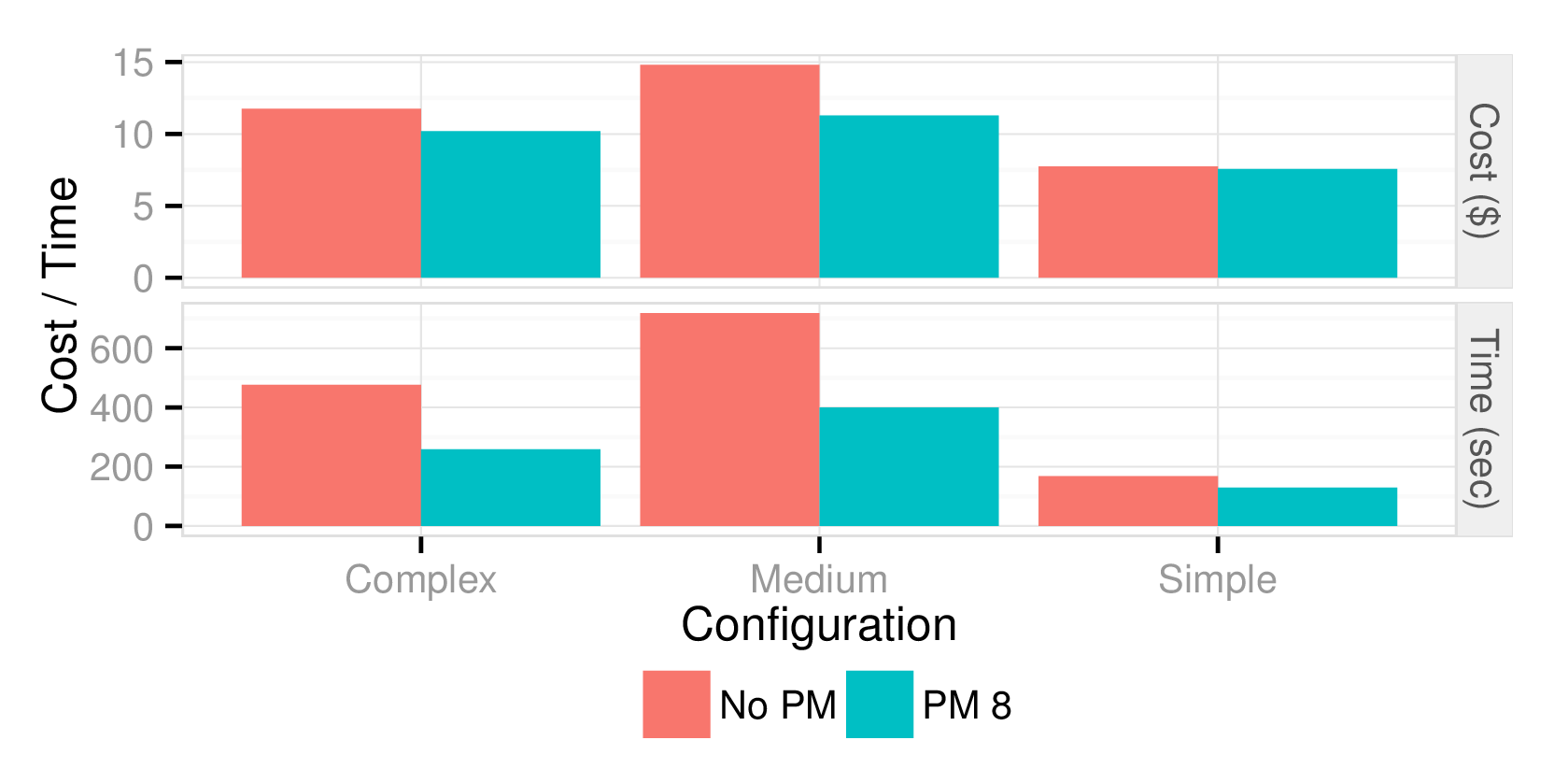}
\end{center}
\vspace*{-.1in}
\caption{Summary of end-to-end cost and latency experiments with and without pool maintenance.}
\label{fig:churn-summary}
\vspace*{-.1in}
\end{figure}

\paragraph{Overall.} Ultimately, pool maintenance does not improve end-to-end latency for simple tasks significantly, but is able to reduce the latency for medium and complex tasks by $1.3\times$ and $1.8\times$ on average, respectively (Figure~\ref{fig:churn-summary}).
Interestingly, despite its added cost to recruit workers concurrently with labeling tasks, maintenance is able to reduce the overall cost of the medium and complex tasks by $7 - 16\%$.  
This is due to finishing the experiment faster and saving the cost of paying workers to stay in the retainer pool.
Changing the rate paid to waiting workers may increase or reduce this effect.

\begin{figure}[h]
\vspace*{-.15in}
\begin{center}
\includegraphics[width=\columnwidth]{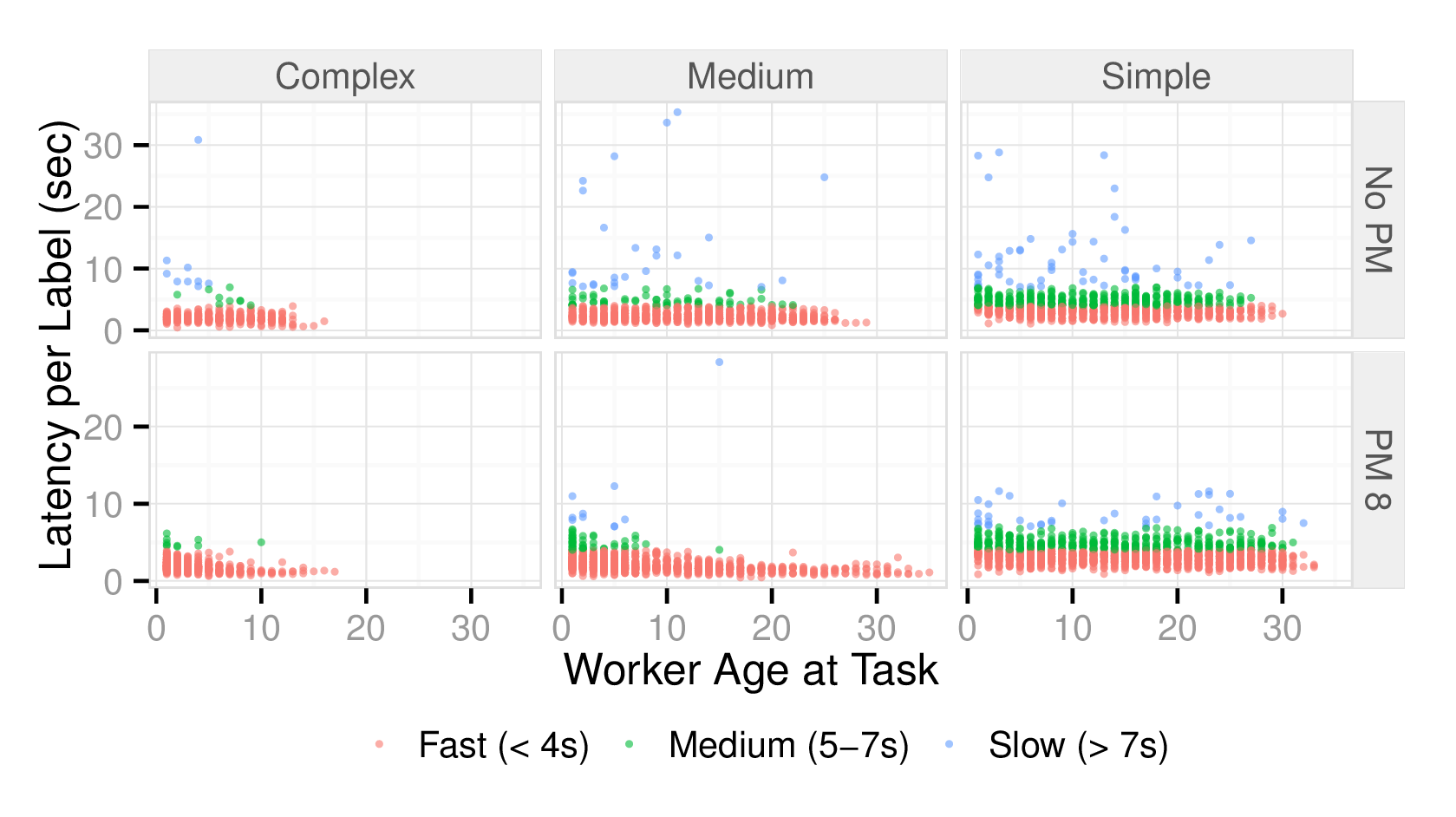}
\end{center}
\vspace*{-.15in}
\caption{Comparison between age of the worker in the pool when starting a given task and the time to complete the task. Tasks where the latency per labeled point is greater than $8$ seconds are colored in blue.}
\label{fig:churn-age-latency}
\vspace*{-.2in}
\end{figure}

\paragraph{Latency Distribution.} To better understand how pool maintenance effects the composition of the worker pool, Figure~\ref{fig:churn-age-latency} plots 
task completion speeds against the age of the worker when starting a given task.  
We define a worker's age with respect to task $t_i$ as the number of tasks the user has already completed in the experimental run.
The y-axis shows the latency to acquire a single label, computed as $\frac{\textrm{task latency}}{N_g}$;
each column shows all tasks across the runs for a given task complexity; and the top and bottom rows are with maintenance turned on ($P\!M_8$) and off ($P\!M_{\infty}$).
In addition, the points are categorized as fast ($<4$ sec per label), medium ($5-7$ sec), or slow ($\ge 8$ sec).
Although workers that are new to the worker pool naturally exhibit high task latency variability, maintenance is able to purge the slow workers over time.  
For every task complexity, the slow and even medium latency tasks are nearly all removed once workers have remained in the pool for more than $4$ minutes.  In contrast, the lack of pool maintenance allows slow and highly variable workers continue working on tasks, so that slow tasks are seen throughout the entire experiment.

\begin{figure}[h]
\vspace*{-.15in}
\begin{center}
\includegraphics[width=\columnwidth]{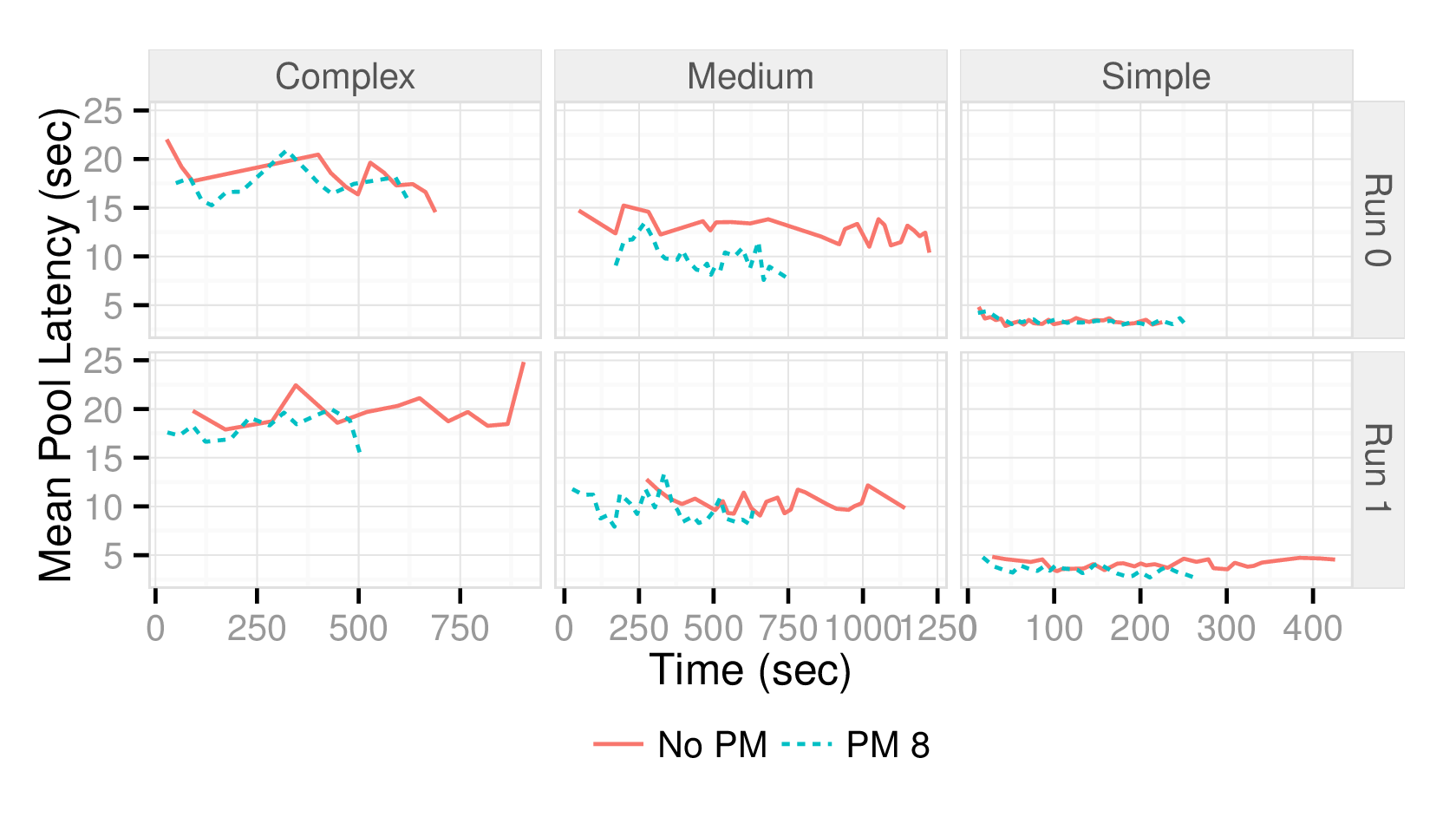}
\end{center}
\vspace*{-.2in}
\caption{Mean pool latency over time.}
\label{fig:churn-mpl}
\vspace*{-.2in}
\end{figure}

\paragraph{Mean Pool Latency.} Figure~\ref{fig:churn-mpl} provides a different view on pool maintenance's effects on the worker pool -- it measures the \textit{mean pool latency (MPL)} for each batch of tasks sent to the pool throughout the experiment.   
MPL is computed as the average latency of all completed tasks in the pool.
Each subplot compares the MPL with and without maintenance for a given experimental run and task complexity.
While the average of each pair of curves is similar, pool maintenance shows significantly less variance across the batches because it effectively removes the long tail of the latency distribution.  
The variation in the pool maintenance curve is simply due to the variation of the newly recruited workers.

\begin{figure}[h]
\vspace*{-.2in}
\begin{center}
\includegraphics[width=\columnwidth]{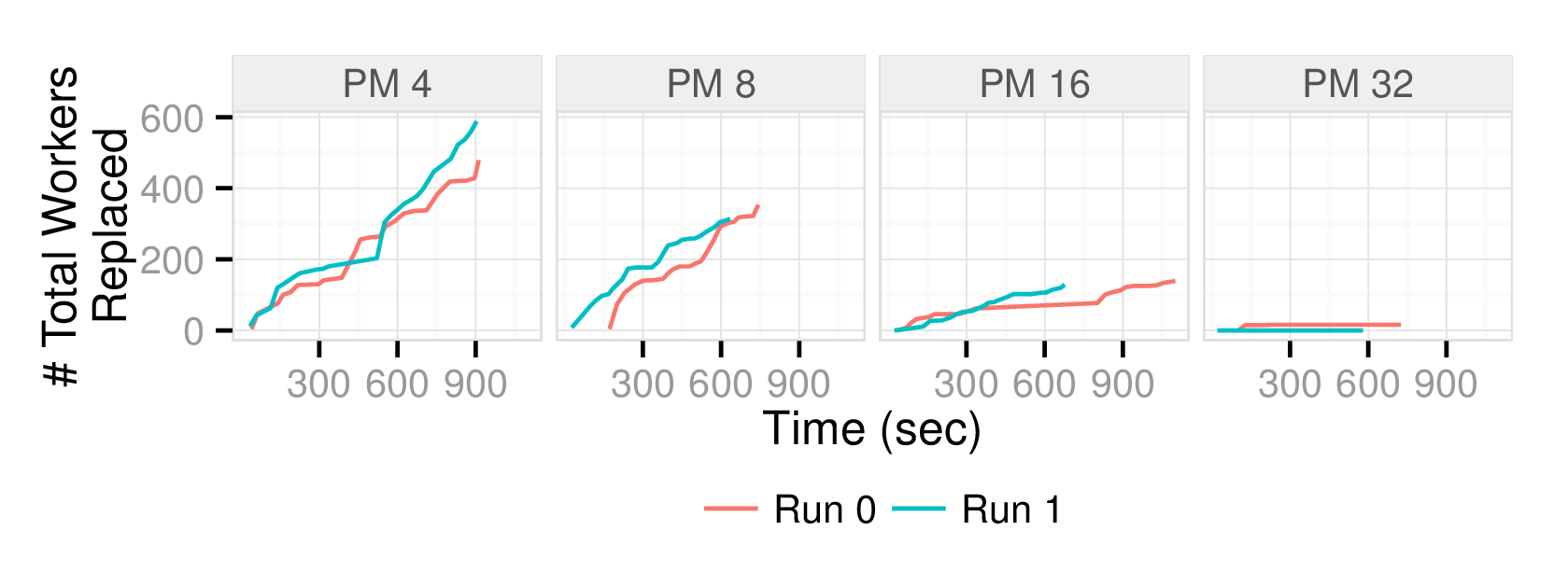}
\end{center}
\vspace*{-.2in}
\caption{The number of workers replaced over time for varying maintenance latency thresholds.}
\label{fig:nchurns}
\vspace*{-.2in}
\end{figure}

\begin{figure}[h]
\vspace*{-.1in}
\begin{center}
\includegraphics[width=\columnwidth]{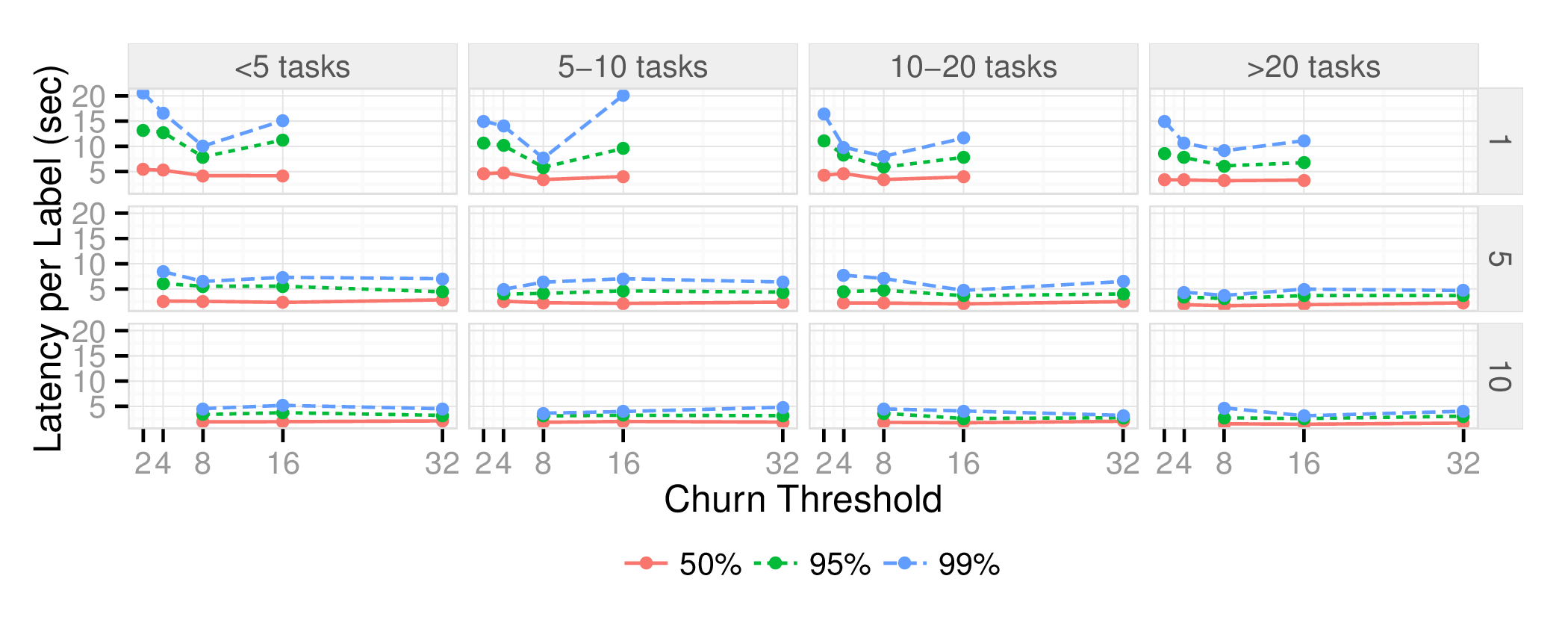}
\end{center}
\vspace*{-.2in}
\caption{$50^{th}$, $95^{th}$, and $99^{th}$ percentiles of task latency as maintenance latency threshold varies.  Each facet is a different amount of time into the experiment.}
\label{fig:churn-threshold}
\vspace*{-.1in}
\end{figure}

\paragraph{Latency Threshold.} Our analysis of MPL shows that pool maintenance is able to remove outliers from the worker pool.
However, the reduction in MPL is not as fast as predicted by the model or simulations presented in Section~\ref{sec:churn}.
This is expected, as workers may not maintain consistent speed over time, and our empirical estimates of worker's speed may be inaccurate.
Another potential issue may be that our latency threshold is poorly tuned, thus in our final experiment (Figures~\ref{fig:nchurns}~and~\ref{fig:churn-threshold}), we study whether varying the latency threshold between $2$ and $32$ seconds can affect the median task latency in addition to the variance. 
Figure~\ref{fig:nchurns} demonstrates that decreasing the threshold causes more workers to be replaced during a run, as expected.
Figure~\ref{fig:churn-threshold} shows the latency percentiles at different worker-age slices (e.g., $<5$ tasks) in the experiment.
We find that varying the threshold affects both the median and higher percentiles, with a more pronounced effect on the extrema task latencies.  
For this workload, the optimal threshold is $P\!M_8$, which can reduce the straggler latencies by nearly $2\times$.
However, further reducing the threshold to $4$ or $2$ seconds goes beyond the point where even fast workers are able to complete tasks, and effectively
replaces all workers with the mean of the underlying MTurk distribution.  The curves reduce across work slices due to the effects of pool maintenance, consistent with the analysis in Figure~\ref{fig:churn-age-latency}.

\vspace{1.0cm}
\subsection{Straggler Mitigation}
\label{sec:exp-straggler}

In this section, we evaluate the performance of straggler mitigation along two key metrics: task latency and task variance.
An important parameter of straggler mitigation is $R$, the ratio of workers in the pool to tasks in a batch (Table~\ref{tab:exp-params}), because it controls how many workers are assigned on average to eliminate stragglers.
Set too low, and stragglers will occur unfettered.
Set too high, and money and effort will be wasted unnecessarily.
In these experiments, we set task complexity to $N_g = 5$, the pool size to $N_p = 15$, and give workers CIFAR-10 tasks.

\begin{figure}[]
\vspace{-.1in}
\begin{center}
\includegraphics[width=\columnwidth]{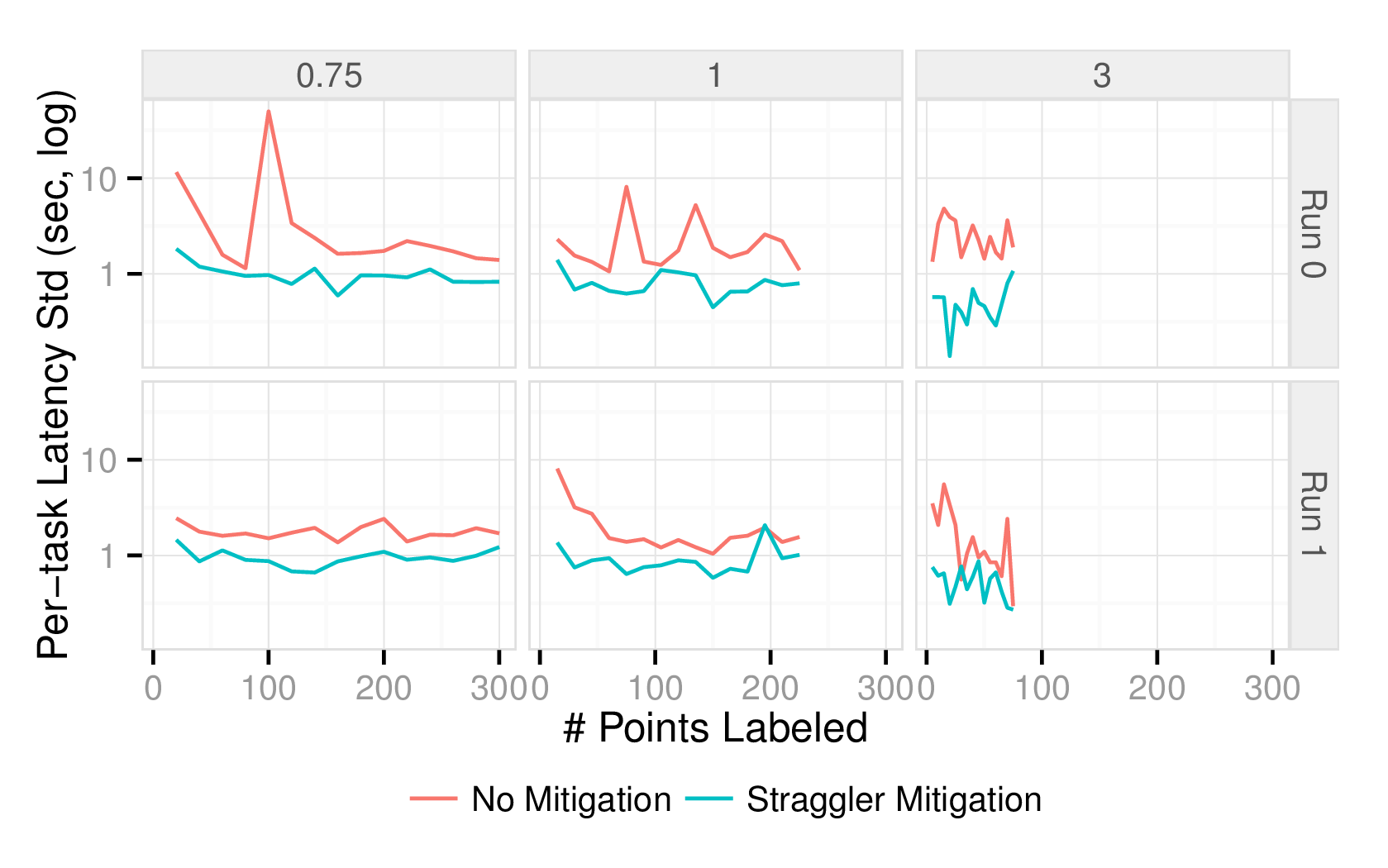}
\end{center}
\vspace{-.2in}
\caption{Straggler mitigation dramatically reduces the standard deviation of per-task latency across batches.}
\vspace{-.1in}
\label{fig:strag-std}
\end{figure}

\paragraph{Variance.} One of the key properties of straggler mitigation is its ability to reduce the variance of individual task latencies.
Figure~\ref{fig:strag-std} plots the standard deviation of the latencies of task completion times for each batch.
Straggler mitigation consistently decreases the standard deviation by 5 to $10\times$ (a decrease in variance of up to $100\times$!), very important when trying to predict the run-time of a batch consistently.
One interesting observation is the jaggedness of the $R=3$ plots. 
This is likely because with 3 times as many workers as tasks, workers spend much more time waiting, and are slow to respond when work becomes available because they are involved in other work.

\begin{figure}[]
\vspace*{-.1in}
\begin{center}
\includegraphics[width=\columnwidth]{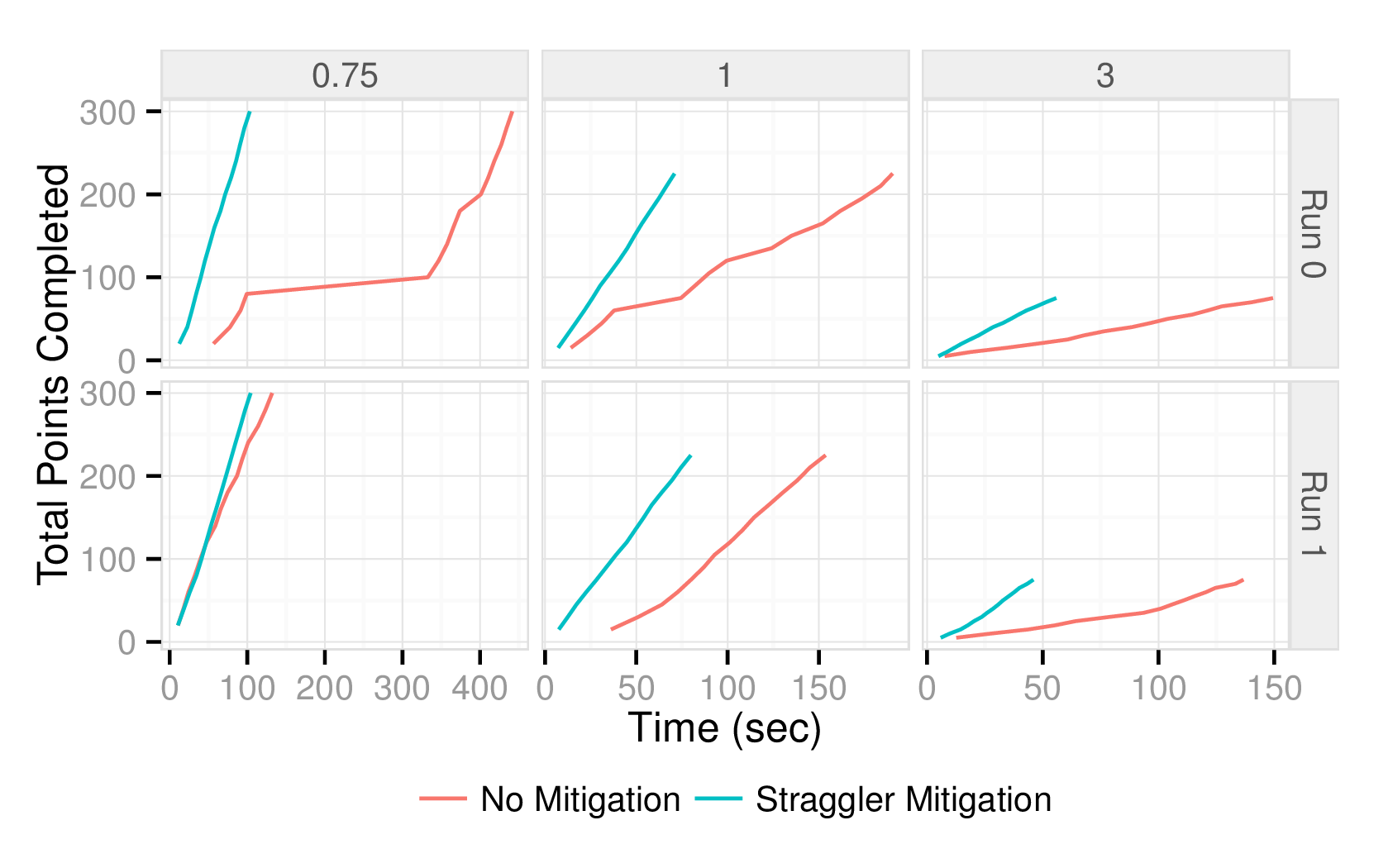}
\end{center}
\vspace*{-.1in}
\caption{Points labeled over time with straggler mitigation}
\label{fig:strag-latency}
\vspace*{-.1in}
\end{figure}

\paragraph{Latency.}
Because straggler mitigation enables task batches to finish without waiting for high-latency straggler task assignments to complete, it significantly reduces the latency of each batch, up to $5\times$ on some runs (Figure~\ref{fig:strag-latency}).
Increasing $R$ can increase those gains, but comes at an additional cost, as it pays more workers to complete each task.
Although intuitively we might expect straggler mitigation to become more and more effective as $R$ increases, there are practical limitations that prevent this effect.
With high $R$, even fast workers are often terminated before finishing their tasks because many workers are working on every task at once.
In addition to the added latency of this termination (workers must click a dialog to finish the old task and be presented with a new one, which takes seconds), this creates a frustrating environment for workers, who feel as though they aren't being allowed to work.
As a result, keeping $R$ between 0.75 and 1 is attractive, as it limits cost and still shows impressive speedups.

\begin{figure}[h]
\vspace{-.1in}
\begin{center}
\includegraphics[width=.8\columnwidth]{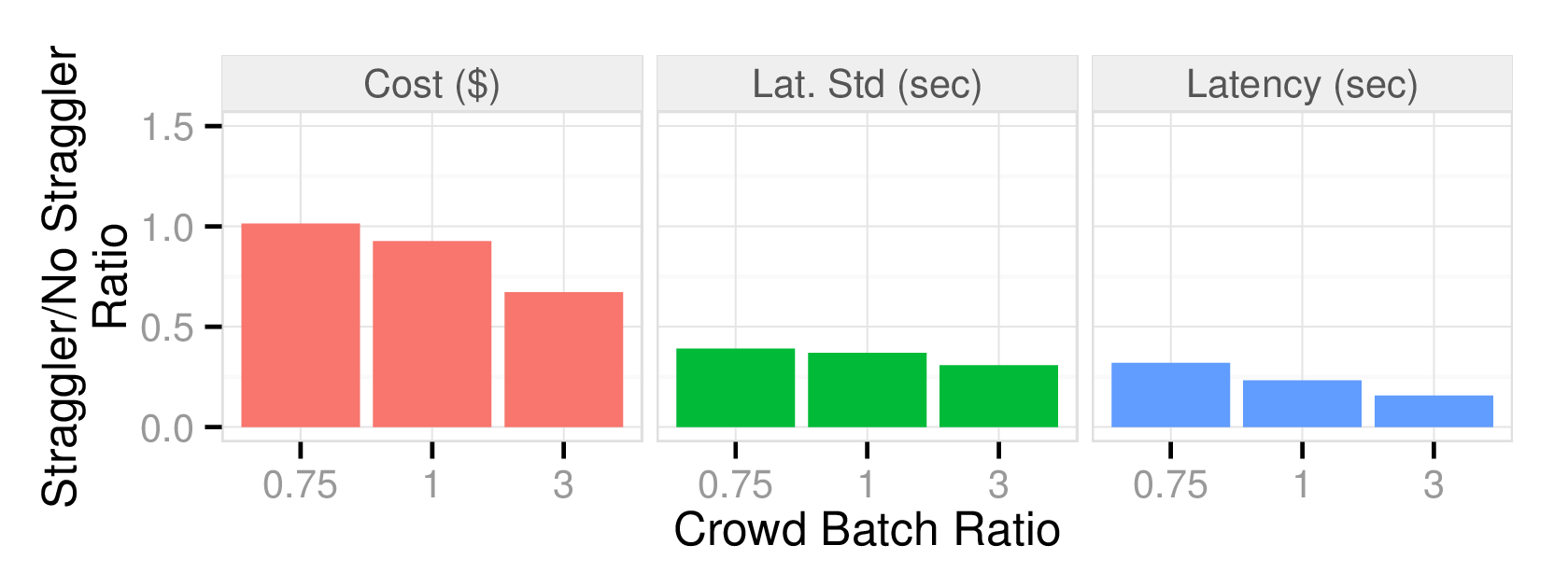}
\end{center}
\vspace{-.2in}
\caption{Straggler mitigation increases costs by 1 to 2$\times$, improves latency by $2.5-5\times$, and variance by $4-14\times$.}
\vspace{-.1in}
\label{fig:strag-ratio}
\end{figure}

\subsection{Combining Per-Batch Techniques}
\label{s:exp-strag-churn}

\begin{figure}[]
\vspace*{-.1in}
\begin{center}
\includegraphics[width=\columnwidth]{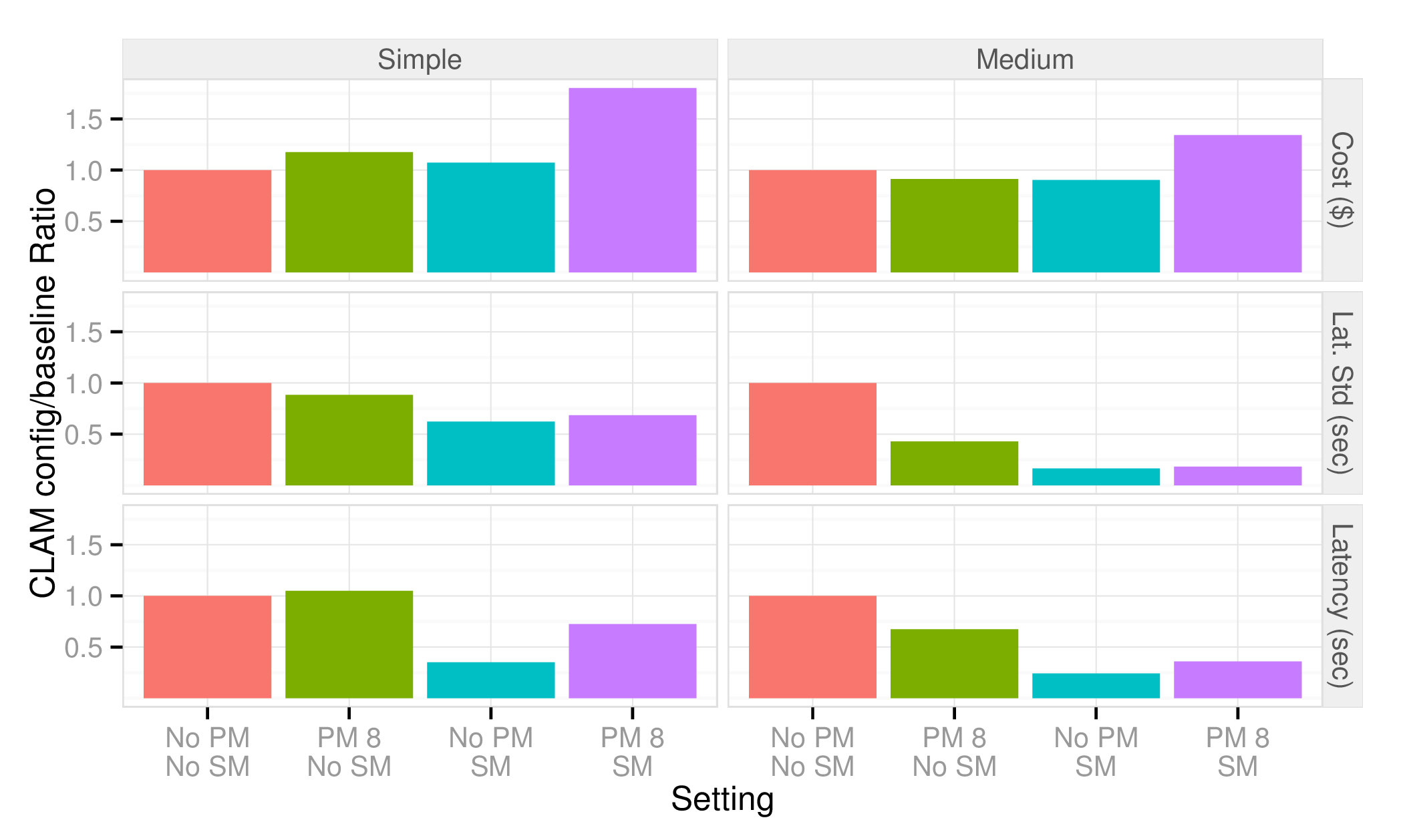}
\end{center}
\vspace*{-.1in}
\caption{End-to-end Latency, Variance, and Costs for different straggler mitigation and pool maintenance configurations.}
\label{fig:stragchurn-summary}
\vspace*{-.1in}
\end{figure}

Figure~\ref{fig:stragchurn-summary} summarizes the effects of combining both straggler mitigation and pool maintenance when labeling CIFAR-10 tasks.
We see that the two techniques can be complementary, but in some experiments we observe destructive interference between straggler mitigation and pool maintenance.
We believe this is a result of fluctuating conditions on the underlying crowd platform across experiments: sometimes the initial pool selection is high-quality, rendering pool maintenance ineffective, and other times very slow workers join the pool and maintenance is invaluable.
We note that in all cases, combining per-batch techniques still results in a significant speedup over not using either technique,
leading to a reduction in latency of up to $6\times$, and reduction in standard deviation of up to $15\times$.

\begin{figure*}[t]
\vspace*{-.1in}
\begin{center}
\includegraphics[width=.8\textwidth]{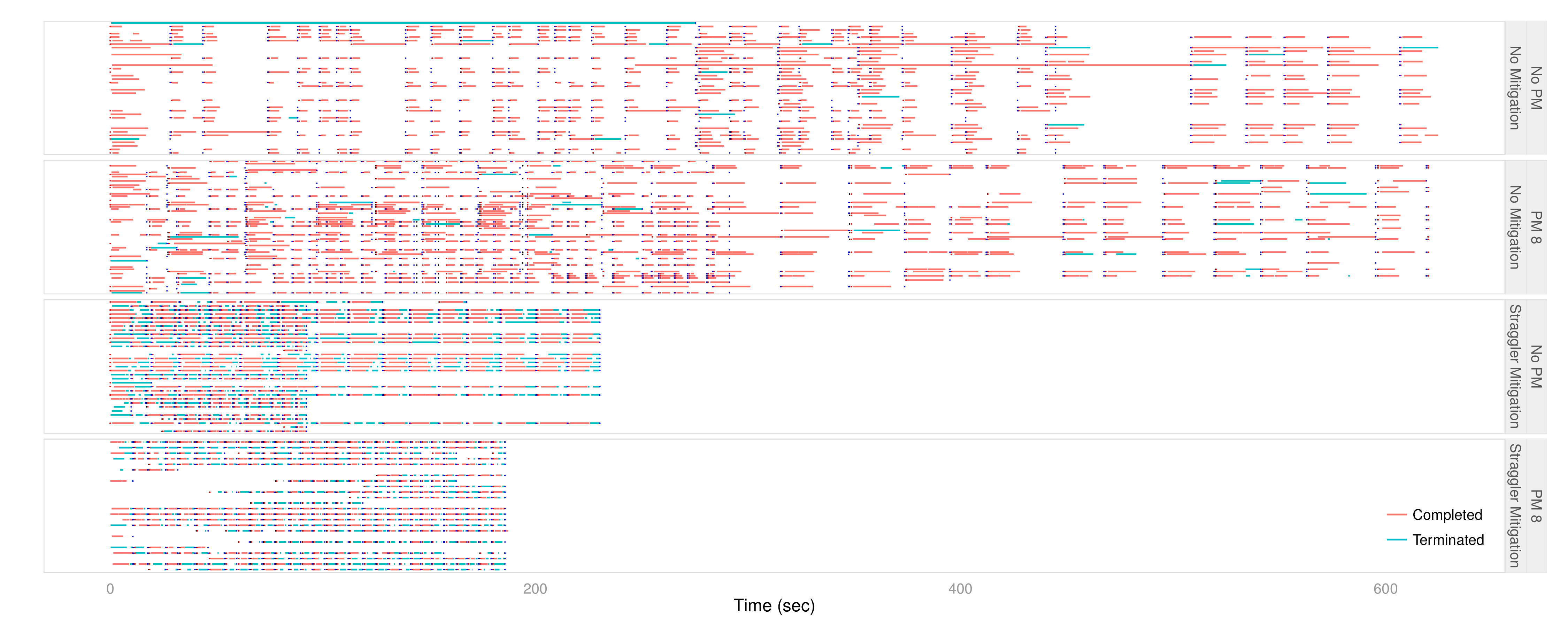}
\end{center}
\vspace*{-.1in}
\caption{Per-assignment view of each straggler mitigation and churn configuration.  Each horizontal segment is the length of an assignment. Red and blue dots denote batch boundaries}
\label{fig:sc-tasklen}
\vspace*{-.1in}
\end{figure*}

\paragraph{Detailed View.} Figure~\ref{fig:sc-tasklen} shows the latency of every task for a single experimental run with every combination of straggler mitigation and pool maintenance.  
Each line segment depicts the start and end time of a specific task.
Red tasks are successfully completed, while blue tasks are terminated due to the worker leaving the pool or because another worker finished the task in less time.  
Red and blue dots denote the start and end of a batch, and the tasks completed by a given worker are aligned vertically along the y-axis.

The top two subplots show the value of pool maintenance -- although stragglers are still present under pool maintenance, there are considerably fewer and lower magnitude stragglers as compared to the baseline pool.  The bottom two subplots show that maintenance can further improve straggler mitigation by reducing the number of stragglers that must be ameliorated.

\begin{figure}[h]
\vspace*{-.1in}
\begin{center}
\includegraphics[width=\columnwidth]{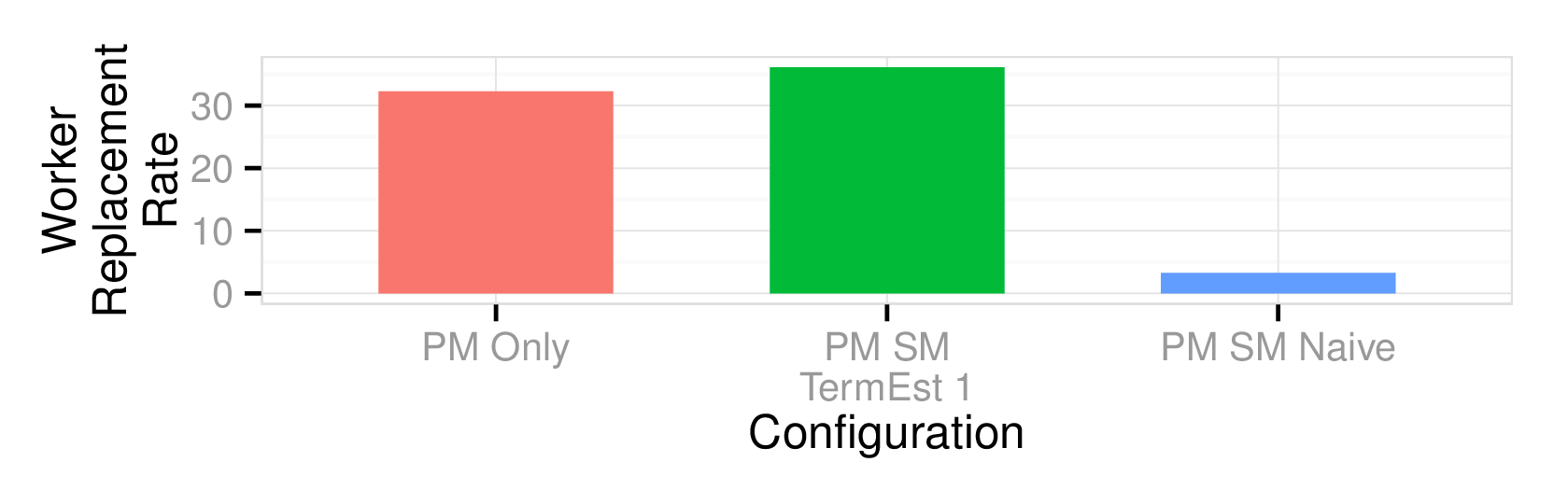}
\end{center}
\vspace*{-.1in}
\caption{Replacement rate when using \texttt{TermEst} (Section~\ref{sec:stragchurn}) with $\alpha=1$.}
\label{fig:termest-nchurns}
\vspace*{-.1in}
\end{figure}

\paragraph{Effect of \texttt{TermEst}.} Figure~\ref{fig:termest-nchurns} measures the effectiveness of our model for estimating the latency of terminated tasks (Section~\ref{sec:stragchurn}).
We see that, as expected, without \texttt{TermEst}, the worker replacement rate decreases dramatically, because workers are estimated to be faster than $P\!M_\ell$ and are not replaced.
Adding \texttt{TermEst} adjusts for the gap: with it turned on, replacement happens just as frequently as with no straggler mitigation.

\subsection{Hybrid Learning}
\label{sec:exp-hybrid}
In this section, we evaluate our hybrid learning strategy, demonstrating that it is effective on datasets where either active or passive learning would perform better, and that it successfully takes advantage of pool parallelism to reduce the time required to train a good model.

\begin{figure}[t]
\vspace{-.1in}
\begin{center}
\includegraphics[width=\columnwidth]{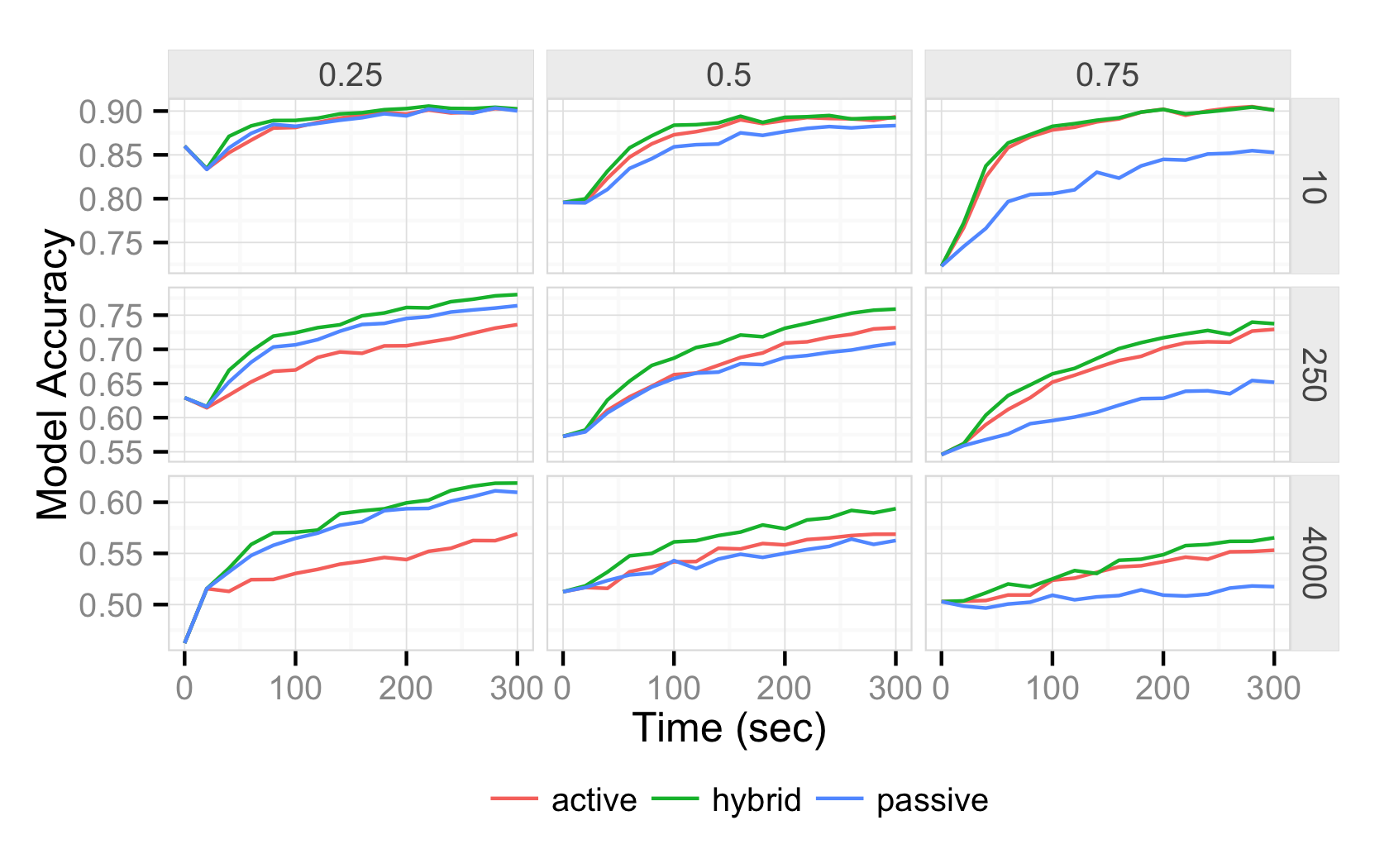}
\end{center}
\vspace{-.2in}
\caption{Active, Passive, and Hybrid strategies for learning on crowds run on generated datasets in the simulator.}
\vspace{-.2in}
\label{fig:hybrid-al-pl-sim}
\end{figure}

\begin{figure}[t]
\vspace{-.1in}
\begin{center}
\includegraphics[width=\columnwidth]{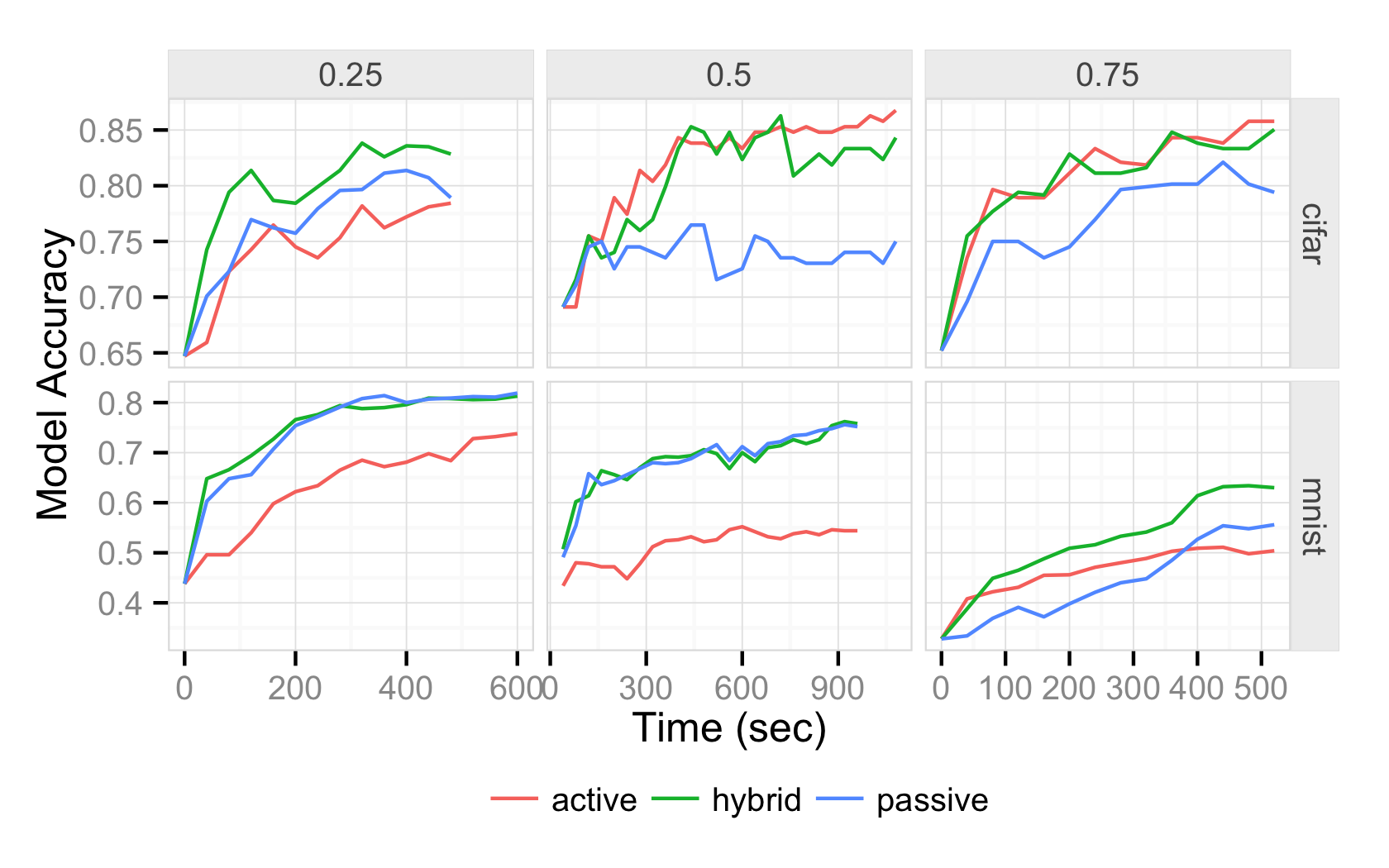}
\end{center}
\vspace{-.2in}
\caption{Active, Passive, and Hybrid strategies for learning on crowds run on real-world datasets on live workers.}
\vspace{-.2in}
\label{fig:hybrid-al-pl}
\end{figure}

\paragraph{Accuracy.} The Hybrid algorithm depends on the assumption that active learning does not outperform passive learning in all settings.
Figure~\ref{fig:hybrid-al-pl-sim} validates this assumption in our simulator.
It plots learning curves for active and passive learning on generated datasets of increasing hardness (rows show number of generated features), and shows how each learner performs given different amounts of the crowd's resources (columns show the percentage of the crowd pool used for active learning).
On easier datasets, active learning significantly outperforms passive learning, but when given as many resources as active learning, passive learning is the better choice on harder learning tasks where active point selection is ineffective.
This reinforces our belief that a successful hybrid strategy can trade off between the two approaches, and the hybrid lines in both Figure~\ref{fig:hybrid-al-pl-sim} and Figure~\ref{fig:hybrid-al-pl} (wherein we replicate the simulator results on real-world datasets with live workers) demonstrate that the strategy is indeed successful. 
In all cases, hybrid performs as well as or better than either active or passive learning.

\paragraph{Latency savings.} As a result of the fact that hybrid learning leverages the full parallelism of the crowd (as opposed to active learning with a limited batch size), the hybrid learning strategy is able to train better models faster. 
Figure~\ref{fig:hybrid-al-pl} shows the hybrid learning strategy's performance compared to pure active or pure passive over time on the MNIST and CIFAR datasets.
The x and y axes of each plot show the accuracy improvement over time as points are labeled, the rows depict the datasets, and the columns represent the setting of the AL batch size as a percentage of the crowd pool size.
In the same amount of time, the hybrid strategy is always the preferred solution for model training.
In fact, on average, hybrid trains models of 85\% accuracy on CIFAR (70\% accuracy on MNIST) $1.2\times$ ($1.7\times$) faster than pure active learning and $1.6\times$ ($1.2\times$) faster than pure passive learning.

\subsection{End-to-End Evaluation}
In this section, we evaluate the end-to-end performance of \sys against two baselines.
\texttt{Base-NR}, which represents a typical crowd labeling deployment, sends labels out all at once, uses no retainer pool, and trains passive learning models to infer labels for unlabeled records.
\texttt{Base-R}, which leverages the latest techniques for low-latency crowdsourcing, uses a retainer pool to label points in batches and active learning to infer labels for unlabeled records.
In this experiment, 500 points were labeled by each strategy on the CIFAR-10 and MNIST datasets, and the accuracy of the resulting models were measured.

\begin{figure}[h]
\vspace*{-.1in}
\begin{center}
\includegraphics[width=\columnwidth]{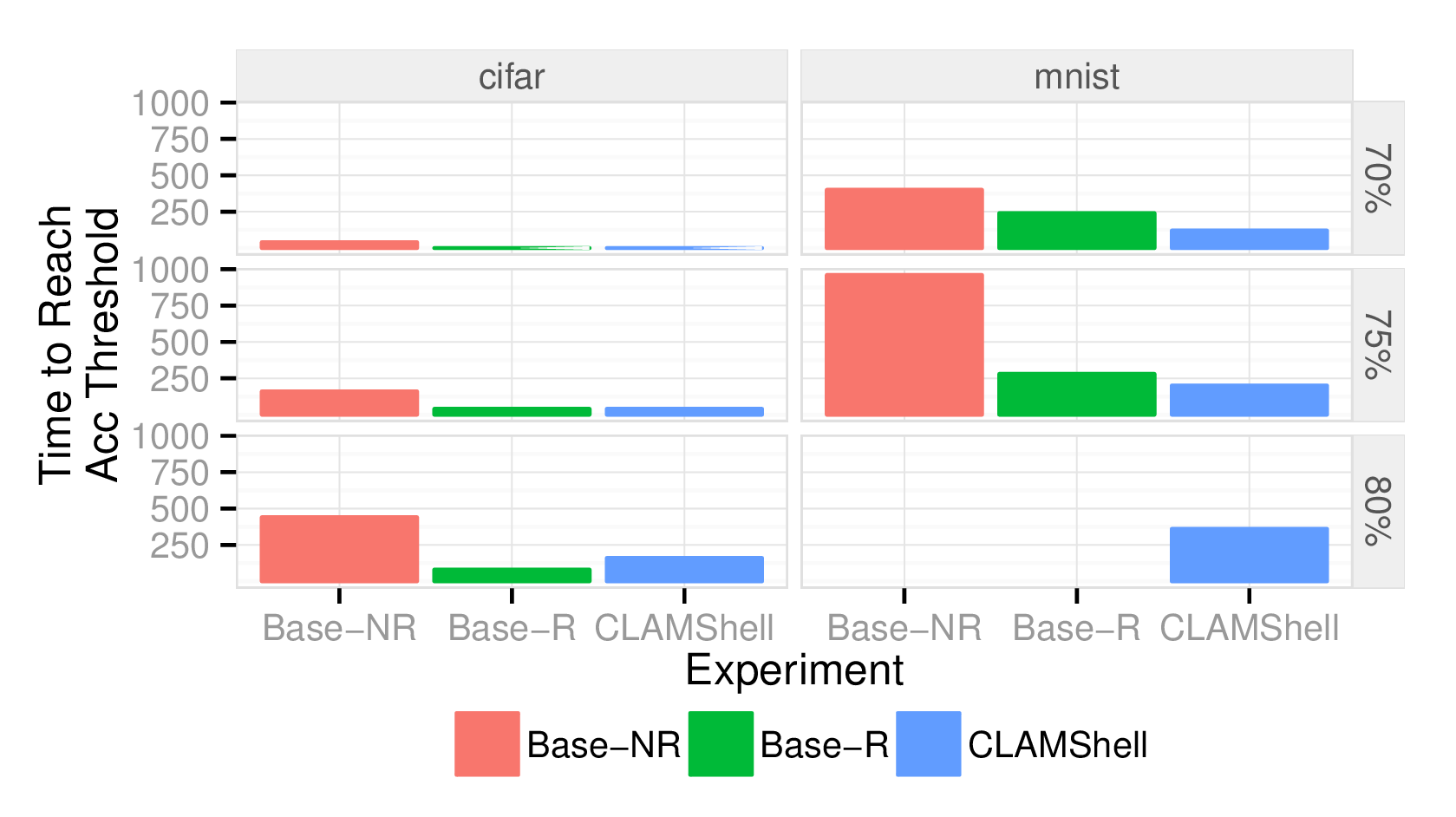}
\end{center}
\vspace*{-.1in}
\caption{Summary of end to end to reach model accuracy}
\label{fig:endtoend-summary}
\vspace*{-.1in}
\end{figure}

\begin{figure}[h]
\vspace*{-.1in}
\begin{center}
\includegraphics[width=\columnwidth]{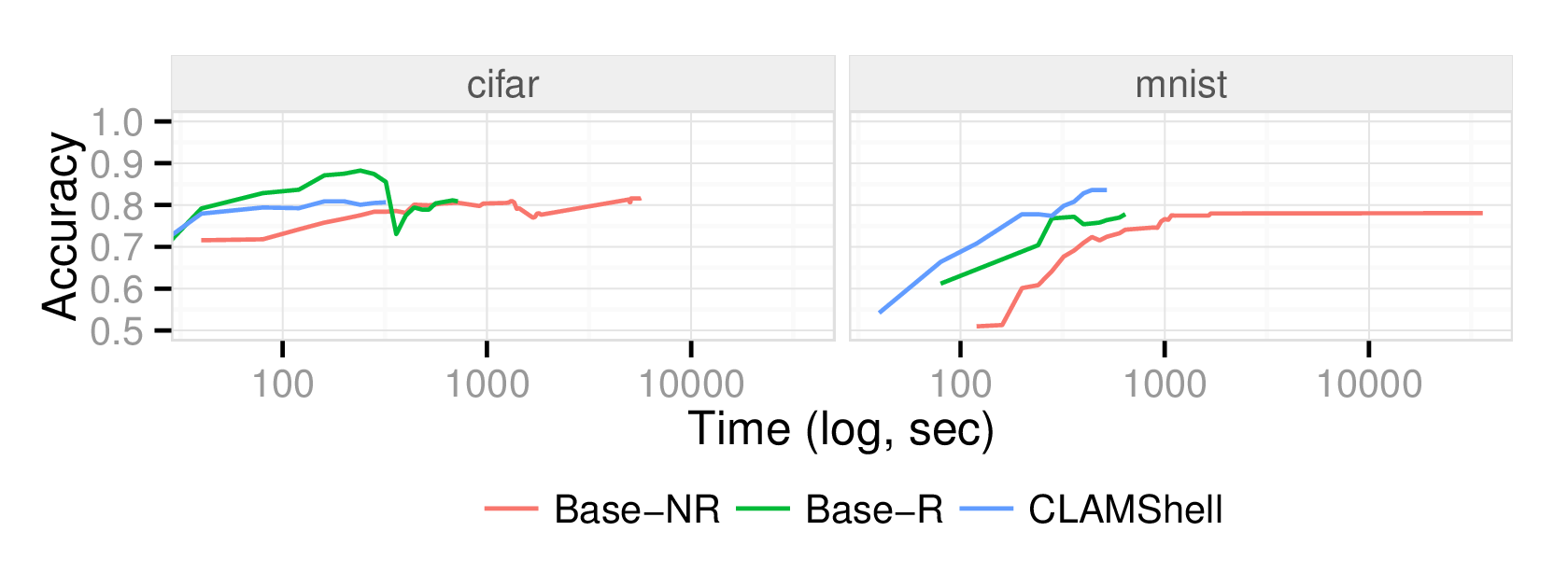}
\end{center}
\vspace*{-.2in}
\caption{Wall clock time vs Model Accuracy}
\label{fig:endtoend-time}
\vspace*{-.2in}
\end{figure}

\paragraph{Results.} Figures~\ref{fig:endtoend-summary} and~\ref{fig:endtoend-time} summarize the results of this evaluation. In Figure~\ref{fig:endtoend-summary}, the rows represent an accuracy threshold for the model, and the plots show the wall-clock time taken by each strategy to train a model of that accuracy. 
Note that neither baselines reach an accuracy of 80\% on the MINST dataset in 500 points.  To reach an accuracy of 75\%, \sys requires $4$ to $5\times$ less time than \texttt{Base-NR}. Figure~\ref{fig:endtoend-time} displays the full learning curves for each strategy, demonstrating that \sys dominates both baselines in terms of model accuracy.

We also measured the raw time to acquire 500 labels from the crowd, and found that \sys increases the labeling throughput by $7.24\times$ compared to \texttt{Base-NR}. In addition, \sys reduces the variance of labeling by $151\times$, and the absolute values are extremely low: 3.1 seconds vs. 475 seconds.

\section{\!\!\!Conclusion \& Future Directions}
\label{sec:discussion}

In summary, we have introduced \sys, a system for data labelling that acquires labels from human crowd workers at interactive speeds.
Latency can arise from many points in the labeling lifecycle, and \sys addresses the key sources of latency with novel techniques.
Straggler mitigation reduces the variance of task latencies within a batch by assigning additional workers to complete the task.
Pool maintenance increases the average speed of workers in a labeling pool by replacing slow workers with faster ones over time.
Hybrid learning reduces end-to-end labeling time by combining the fast convergence of active learning with the parallelism of passive learning.
The result is an important step towards integrating data labeling with interactive systems for data analysis.

Though \sys takes a comprehensive approach to latency reduction for data labeling, there are a number of directions in which this work can be extended.
First, we would like to explore richer objective functions than mean worker speed for pool maintenance in order to strike a balance between worker speed, variance and quality.
In addition, hybrid learning simply trains a single model on the points labeled by active and passive learners. 
We would like to investigate whether better models can be trained by keeping the points separate and using more sophisticated machine learning techniques such as model averaging or ensembling.
Finally, we are integrating \sys with an interactive data cleaning system~\cite{Haas:2015a} in order to learn how it performs with application-driven latency constraints on a wider range of crowd tasks.

\newcommand{\refname}{\normalfont\selectfont\normalsize References} 
\fontsize{7pt}{7pt}\selectfont

\balancecolumns
\end{document}